\documentclass[journal]{IEEEtran}
\usepackage{amsmath,amsfonts}
\usepackage{algorithmic}
\usepackage{booktabs}
\usepackage{comment}
\usepackage{multirow}
\usepackage{algorithm}
\usepackage{array} 
\usepackage{textcomp}
\usepackage{stfloats}
\usepackage{cite}
\usepackage{url}
\usepackage{verbatim}
\usepackage{times}
\usepackage{epsfig}
\usepackage{graphicx}
\usepackage{newtxmath}
\usepackage{xcolor}
\usepackage[pagebackref,breaklinks,colorlinks,
            linkcolor=red,
            anchorcolor=red,
            urlcolor= blue,
            citecolor=blue]{hyperref}

\begin{document}
\title{A Review of Uncertainty Estimation and its Application in Medical Imaging} 

\author{Ke Zou, Zhihao Chen, Xuedong Yuan, Xiaojing Shen, Meng Wang, and Huazhu Fu,
\thanks{This work was supported by Key Research and Development Program of Sichuan Province (Grant No. 2023YFG0273) and Miaozi Project in the Science and Technology Innovation Program of Sichuan Province (2021001). }
\thanks{K.~Zou and X.~Yuan are with the National Key Laboratory of Fundamental Science on Synthetic Vision, the College of Computer Science, Sichuan University, Chengdu 610065, China. (e-mail: kezou8@gmail.com, yxdongdong@163.com) }
\thanks{Z.~Chen is with the College of Intelligence and Computing, Tianjin University, Tianjin 300350, China. (e-mail: zh\_chen@tju.edu.cn) }
\thanks{X.~Shen is with the Department of Mathematics, Sichuan University, Chengdu 610065, China. (e-mail: shenxj@scu.edu.cn)  }
\thanks{M.~Wang and H.~Fu are with the Institute of High Performance Computing (IHPC), Agency for Science, Technology and Research (A*STAR), Singapore 138632, Republic of Singapore. (e-mail: wangmeng9218@126.com, hzfu@ieee.org)}
\thanks{X.~Yuan and H.~Fu are the co-corresponding authors. }
}


\maketitle

\begin{abstract}
The use of AI systems in healthcare for the early screening of diseases is of great clinical importance. Deep learning has shown great promise in medical imaging, but the reliability and trustworthiness of AI systems limit their deployment in real clinical scenes, where patient safety is at stake.  Uncertainty estimation plays a pivotal role in producing a confidence evaluation along with the prediction of the deep model. This is particularly important in medical imaging, where the uncertainty in the model's predictions can be used to identify areas of concern or to provide additional information to the clinician. In this paper, we review the various types of uncertainty in deep learning, including aleatoric uncertainty and epistemic uncertainty. We further discuss how they can be estimated in medical imaging. More importantly, we review recent advances in deep learning models that incorporate uncertainty estimation in medical imaging. Finally, we discuss the challenges and future directions in uncertainty estimation in deep learning for medical imaging. We hope this review will ignite further interest in the community and provide researchers with an up-to-date reference regarding applications of uncertainty estimation models in medical imaging. 
\end{abstract}


\IEEEpeerreviewmaketitle

\section{Introduction}

\begin{figure}[!t]
\centering
\label{fig_sample}
\includegraphics[width=1\linewidth]{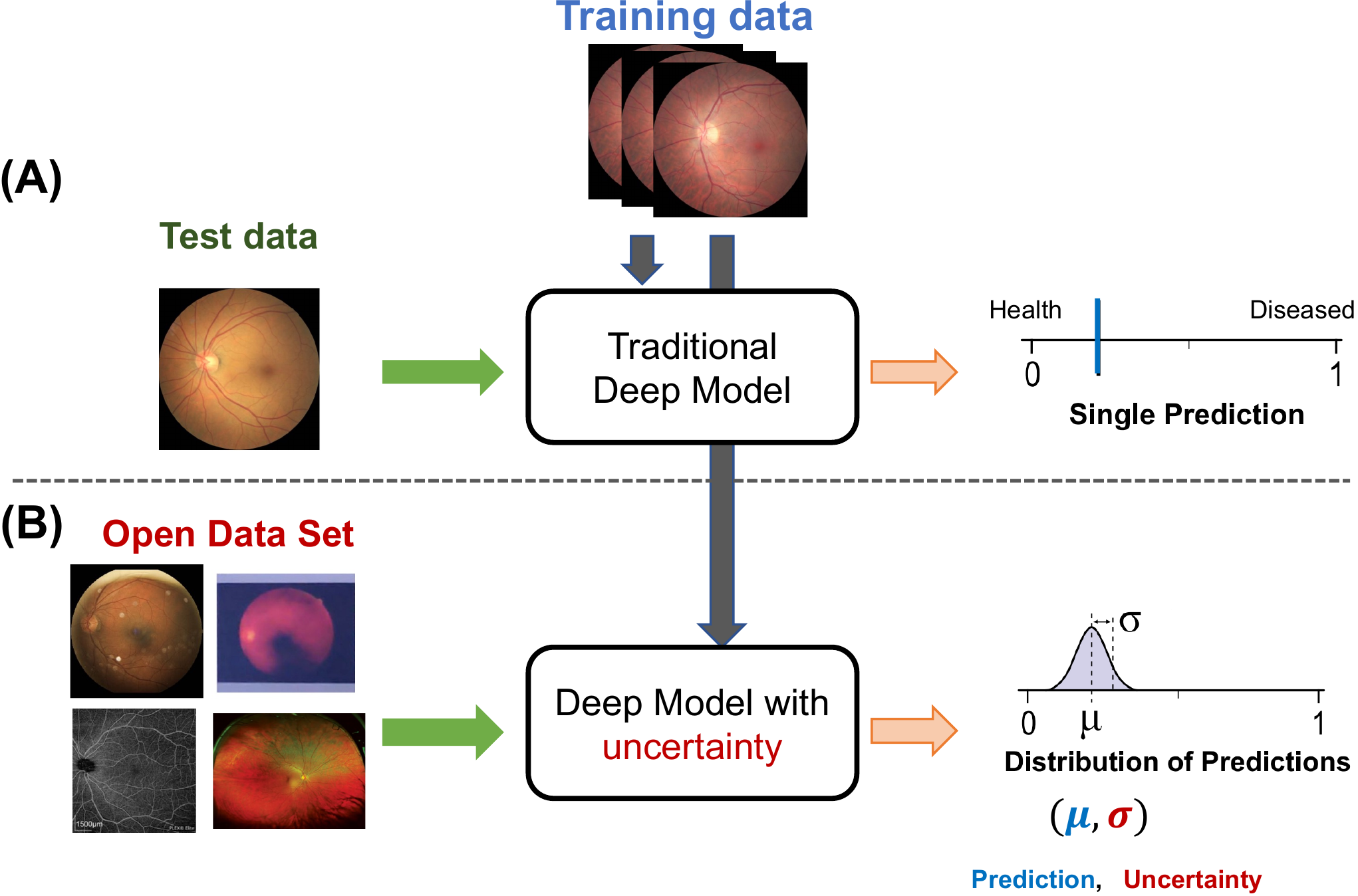} 
\caption{The example of eye disease screening AI system  in the open clinical environment. (A) Traditional deep models are often trained with closed-world assumption, \textit{i.e}, the distribution of test data is assumed to be similar to the training data distribution. (B) However, when deployed in real clinical scenes, this assumption doesn't hold true leading to a significant drop in performance and producing an unreliable result. The uncertainty estimation provides a confidence score, which allows users to quantify the reliability of the model's output and to identify when the model may not be performing well.}
\end{figure}

Deep learning systems achieve significant progress in medical image analysis, and are used widely for a wide range of tasks~\cite{Esteva2017,Lee2019,Bi2019,Rajpurkar2022,Shamshad2022,fundus_survey}, such as tumor segmentation, disease diagnostics, and treatment planning. 
However, these systems can also introduce new risks and challenges, such as bias, errors, and lack of transparency. Imagine a deep learning system that is used to identify eye disease as either Diabetic Retinopathy (DR) or normal. The system has been trained on a large dataset of well-cleaned images of eye diseases collected from hospitals  and is able to make predictions with high accuracy on the test image within the same distribution, as shown in Fig.~\ref{fig_sample} (A). However, there is still some uncertainty in its predictions, as some of the images in the open medical environments with low quality or Out-of-Distribution (OOD) are difficult to classify or contain a high degree of variability. For these low-quality/OOD cases, the transitional AI system may still provide a probability score indicating the likelihood that the eye is diseased. However, this prediction is not reliable. 

Recently, trustworthy AI is proposed to address these challenges by incorporating principles such as explainability, robustness, and accountability into the design and development of AI systems, which intends to provide clinicians and patients with the confidence that the predictions and recommendations made by the AI are accurate and reliable~\cite{Li2022,Liang2022}. \textbf{Uncertainty Estimation} (or Uncertainty Quantification) of deep networks, as one key of trustworthy AI, refers to the process of predicting the uncertainty or confidence of a neural network's predictions~\cite{Kendall2017,Abdar2021,Li2022_CVPR}. Since it allows us to quantify the reliability of the network's output and to identify when the network may not be performing well. For example, when a clinician uses the system to analyze an open set image of the eye disease, the deep learning system still makes predictions with a high degree of accuracy, but it also produces an uncertainty estimation in its predictions, as shown in Fig.~\ref{fig_sample} (B). In this case, the clinician then would be afforded opportunities to take into account the uncertainty in the prediction and consider other factors such as the patient's medical history and any additional diagnostic tests, and either ignore predictions with high uncertainty or triage them for detailed, human review. Therefore, it's important for the clinician to understand this uncertainty and how it may impact the decision making process. 

Uncertainty estimation used in healthcare can be divided into interventional and non-interventional methods according to the doctor's involvement on the data:
\begin{itemize}
\item  Interventional applications: Interventional safety-critical applications involve scenarios where errors can have severe consequences, such as cancer diagnosis. In such case, it is crucial to measure the model's confidence in its predictions. If the model exhibits high uncertainty in its predictions, the results need to be referred to experts for further diagnosis and intervention. This ensures that critical decisions are not solely based on uncertain predictions and that human expertise is involved in the decision-making process.
\item Non-interventional applications: Non-interventional methods do not require expert intervention. They are used for preliminary screening of data, such as out-of-distribution (OOD) sample detection and anomaly detection. \textbf{OOD sample detection} refers to the identification of inputs that significantly differ from the training data, leading to potentially unreliable predictions. By estimating uncertainty, the model can flag such inputs as potentially problematic, triggering further analysis or human intervention. \textbf{Abnormal detection} is another non-interventional application, where the model makes incorrect predictions for abnormal cases that were not present in the training data. Uncertainty estimation helps in identifying instances where the model is likely to be incorrect, highlighting areas for model improvement or the need for additional training data.
\end{itemize}
By employing uncertainty estimation techniques in healthcare, both non-interventional and interventional applications can benefit from improved reliability and safety. The use of uncertainty estimation enhances the decision-making process, reduces risks, and ensures appropriate involvement of medical professionals in critical cases.

\section{Background of Uncertainty}
\subsection{Types of Uncertainty}

 \begin{figure}[!t]
\centering
\includegraphics[width=1\linewidth]{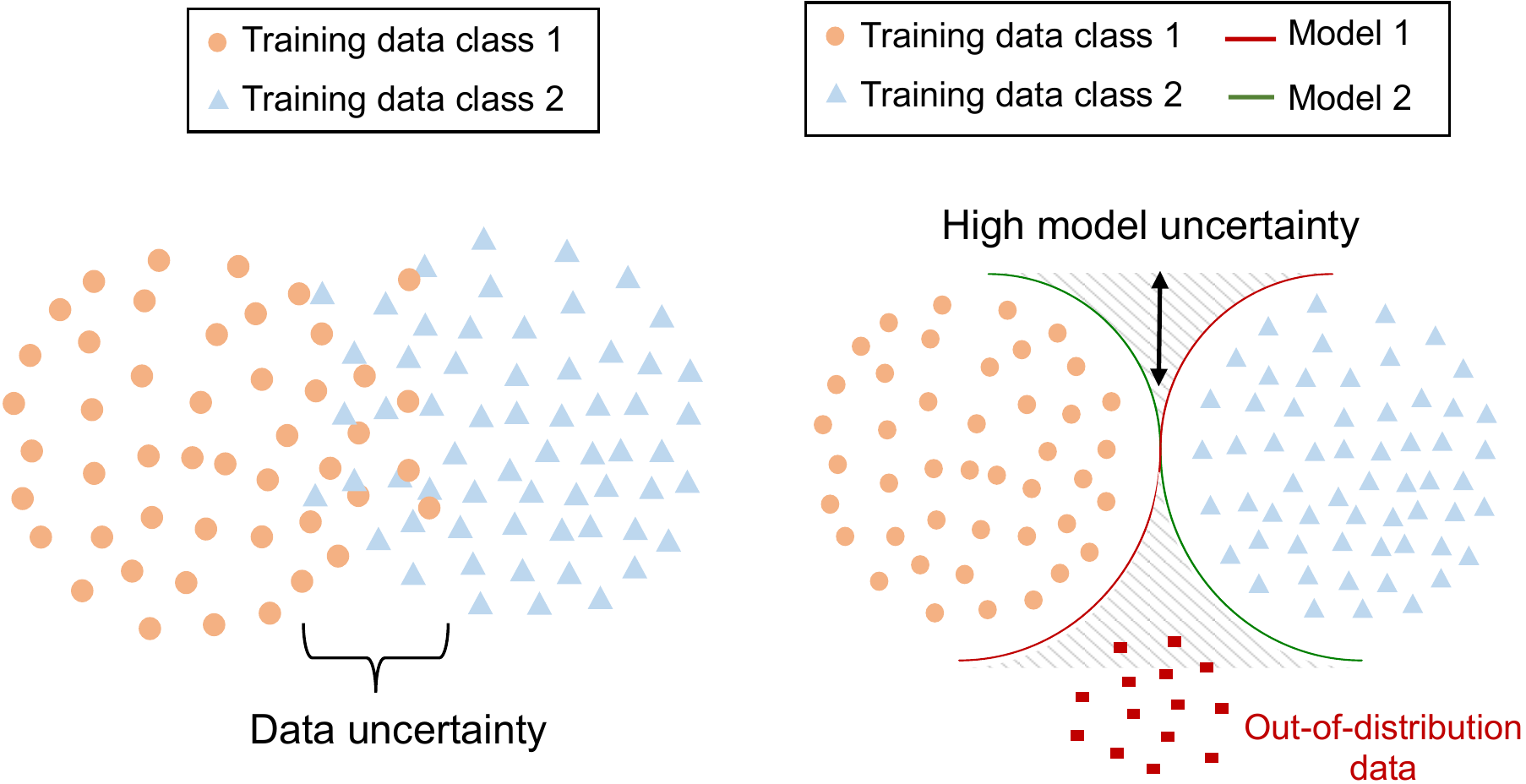}\\
\caption{Visualization of the aleatoric (data) and the epistemic (model) uncertainty for the classification model.}
\label{fig_type}
\end{figure}

Uncertainty in the context of deep learning models refers to the model's lack of confidence in its predictions. This can be thought of as a measure of the model's ignorance or ambiguity about the correct output for a given input. \textbf{There are two main types of uncertainty to quantify in deep learning models~\cite{der2009aleatory}, as shown in Fig.~\ref{fig_type}:}

\textbf{Aleatoric uncertainty (Data uncertainty):} This type of uncertainty arises from inherent noise in the data, such as measurement error or ambiguous annotation, which cannot be reduced by collecting more data~\cite{der2009aleatory,Kendall2017}. Aleatoric uncertainty can be estimated by training the model to output a distribution over possible predictions, rather than a single point estimate. The aleatoric uncertainty will never get smaller, even if we master the problem.

\textbf{Epistemic uncertainty (Model uncertainty):} This type of uncertainty arises from a lack of knowledge or information about the underlying model or data distribution or insufficient model structure~\cite{der2009aleatory}. Epistemic uncertainty can be estimated by using methods such as Bayesian neural networks or ensembles of models. This kind of uncertainty can be (theoretically) reduced  by using more complex models, collecting more data, or by using regularization techniques. In some uncertainty works, they also mention a special kind of uncertainty, \textbf{Distributional uncertainty}, which refers to the uncertainty in the model's predictions when the inputs belong to a different distribution than the training data, i.e., OOD.  In deep learning, this can occur when the model encounters inputs that are significantly different from what it was trained on. The distributional uncertainty could belong to epistemic uncertainty.


\subsection{Methods of Uncertainty Estimation}

\begin{figure*}[!t]
\centering
\includegraphics[width=1\linewidth]{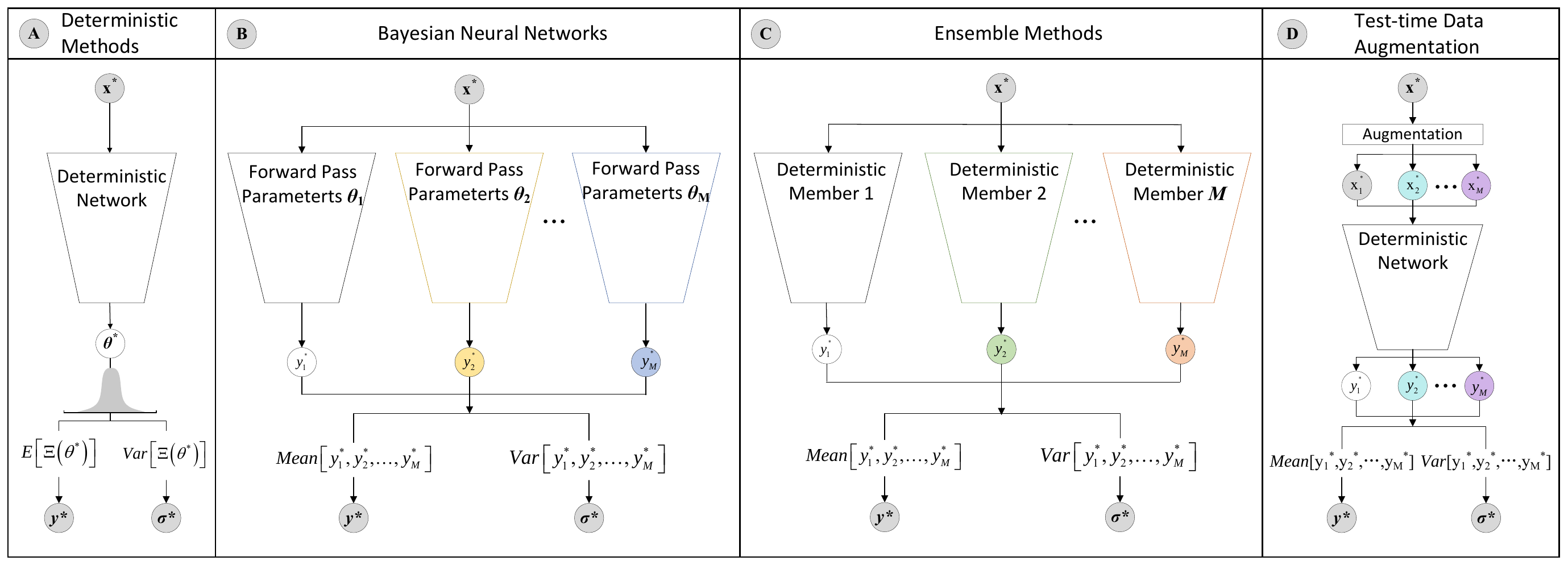}\\
\caption{ The different methods of uncertainty estimation.}
\label{fig_method}
\end{figure*}

There are a few different approaches to estimating uncertainty in deep networks, as shown in Fig.~\ref{fig_method}, including:
\begin{itemize}
\item  \textbf{Deterministic method:} Deterministic methods provide a deterministic estimate of uncertainty, meaning that they provide a single value or measure to represent the level of uncertainty associated with a prediction. These methods typically assume that the model is deterministic, and that the uncertainty can be estimated based on one single forward pass. The main advantage of them is their simplicity and computational efficiency. The common deterministic methods include evidential deep learning~\cite{evidential18,9767662,Han2021_ICLR,Ma2021_NIPS,Zou2023}, and distance-based~\cite{2020NIPSdeterministic, mukhoti2021deep, van2020uncertainty} methods.
\item \textbf{Bayesian Neural Networks (BNNs):} BNNs are a type of deep learning model that explicitly model uncertainty by representing the model's parameters as random variables~\cite{van2017neural,razavi2019generating,maddox2019simple,dera2021premium}. This allows BNNs to estimate uncertainty by quantifying the distribution of possible outputs for a given input, rather than just a single point estimate. An alternative to directly estimating model parameters is to approximate inference from multiple predictions of the model, which saves computational overhead~\cite{Abdar2021}. In this method, Dropout~\cite{srivastava2014dropout} is often used as a regularization technique that involves randomly setting a percentage of the inputs to a layer to zero during training. This can be seen as a way of approximating Bayesian inference by sampling different model architectures during training.
\item \textbf{Ensemble method:} The ensemble method for uncertainty estimation in deep learning involves using multiple models to make predictions, and then aggregating the predictions to estimate uncertainty with the variance of the individual predictions serving as a measure of uncertainty~\cite{ensemble17,2019EnsembleTTA,ashukha2020pitfalls,shen2022DCA}. This can be done by combining the outputs of multiple models, or by training an ensemble of models to make predictions.
\item \textbf{Test-time data augmentation:} It is a method for uncertainty estimation that involves augmenting the test data with different perturbations, and then evaluating the model's performance on these perturbed inputs~\cite{Ezhov2022_miccai,WANG201934,9413010}. This allows the model to estimate its own uncertainty by evaluating how its predictions change for different inputs.
\end{itemize}  
Each of these methods has its own advantages and disadvantages, and the best method will depend on the specific requirements of the task and the application.   

\section{Applications in Medical Imaging}

\subsection{Classification}
The classification task is an important topic in the field of medical image processing. In recent years, with the explosive development of deep learning, numerous methods for medical image target classification have been explored. These methods have indeed achieved remarkable success in medical image classification tasks as well, even achieving results comparable to those of clinicians in some areas. However, most of these approaches have focused on improving the performance of the algorithms and ignored the reliability analysis of the model prediction results, which has become one of the important factors limiting the deployment of AI algorithm-based classification models in clinical practice. Therefore, developing AI models with uncertainty estimation for medical image classification tasks is essential to improve user confidence in deploying AI models as an aid to diagnosis in clinical practice. Depending on the application domain, we will discuss uncertainty estimation for ophthalmology images~\cite{Leibig2017,ARAUJO2020101715, AYHAN2020101724, Jaskari2022}, histopathology images~\cite{belharbi2021deep, dolezal2022uncertainty, feng2022trusted, LINMANS2023102655}, skin disease images~\cite{van2019quantifying, Combalia2020, ABDAR2021104418}, MRI images~\cite{Herzog2020, prince2023uncertainty}, and chest radiographs~\cite{Ghesu2019, irvin2019chexpert, calli2019handling, pham2021interpreting}.

Recently, several studies have been explored to introduce uncertainty theory to conduct trustworthy classification in ophthalmology images. In~\cite{Leibig2017}, Leibig et al. evaluated the dropout based Bayesian uncertainty measure for deep learning in diagnosing DR from fundus images and showed that it captures uncertainty better than direct alternatives. They computed meaningful uncertainty measures without adding additional labels for explicit uncertainty categories based on the connection between dropout networks and approximate Bayesian inference. Besides, they also demonstrated that uncertainty-informed decision referrals can improve diagnostic performance. For DR detection, Araújo et al.~\cite{ARAUJO2020101715} proposed a novel deep learning-based DR grading system, which supports its decision by providing a medically interpretable explanation and an estimation of how confident that prediction is, indicating that the ophthalmologist to measure how much that decision should be trusted. Ayhan et al.~\cite{AYHAN2020101724} introduced an intuitive framework based on test-time data augmentation for quantifying the diagnostic uncertainty of deep neural networks for detecting DR. In addition, based on Bayesian neural networks, Jaskari et al.~\cite{Jaskari2022} further proposed an uncertainty-aware deep learning method for robust DR classification. 

Moreover, inspired by the pathologist’s actual practices and the automatic Whole Slide Image~(WSI) classification system, Feng et al.~\cite{feng2022trusted} proposed a trusted multi-scale classification framework for the WSI based on uncertainty theory. In their study, a vision transformer was employed as the backbone for different branches to model the classification and evidential uncertainty theory was introduced to estimate the uncertainty of each magnification of a microscope. The final classification result is calculated by integrating the evidence from different magnifications. This method achieved excellent performance on two databases: Liver-Kidney-Stomach immunofluorescence WSIs~\cite{maksoud2020sos} and Fibroma hematoxylin-eosin WSIs. Dolezal et al~\cite{dolezal2022uncertainty}. made high-confidence predictions for digital histopathology with an uncertainty-based deep learning model. In this study, they introduced a clinically-oriented approach to uncertainty quantification for whole-slide images, evaluating uncertainty using dropout and calculating thresholds on training data to determine cutoffs for low and high confidence predictions. They trained models to identify lung adenocarcinoma and squamous cell carcinoma and demonstrated that high confidence predictions outperformed uncertain predictions in cross-validation and testing of two large external datasets spanning multiple institutions. Furthermore, Linmans et al~\cite{LINMANS2023102655} explored the introduction of predictive uncertainty estimation to detect OOD detection in digital pathology. This work provides a benchmark for evaluating popular methods on multiple datasets by comparing uncertainty estimates for within-distribution and OOD samples at the whole slide level.

In the field of skin lesion analysis, Molle et al.~\cite{van2019quantifying} first pointed out the limitations of approximating uncertainty inference based on Bayesian estimates and propose a novel uncertainty measure based on overlap of output distributions. And the effectiveness of the metrics was verified in the classification of skin lesion. In addition, Combalia et al.~\cite{Combalia2020} explored the use of uncertainty estimation methods and metrics for deep neural networks and apply MC-Dropout for dermoscopic image classification. More comprehensive, Abdar et al.~\cite{ABDAR2021104418} introduced three uncertainty quantification methods, MC-Dropout, ensemble MC, and deep ensemble to address the uncertainty in skin cancer image classification, as well as proposed a novel hybrid dynamic Bayesian deep learning model that takes uncertainty into account based on the three-branch decision theory. This method achieved encouraging classification performance for skin cancer images.

In the field of Magnetic Resonance Image (MRI) analysis, Herzog et al.~\cite{Herzog2020}. proposed a Bayesian convolutional neural network to predict a probability for a stroke lesion on 2D MR images while generating corresponding uncertainty information about the reliability of the prediction. Prince et al.~\cite{prince2023uncertainty} employed the Variational Inference by elliptical slice sampling to quantify the uncertainty for classification of Adamantinomatous Craniopharyngioma from preoperative MRI. They developed a classification waiver mechanism using uncertainty estimation to support clinical noninvasive diagnosis of brain tumors in the future. In the field of chest radiographs analysis, to address the fact that traditional AI models may have poor generalization to unseen data due to overconfidence in prediction results, Ghesu et al.~\cite{Ghesu2019} proposed an automatic system for chest radiograph assessment based on the principles of information theory and subjective logic~\cite{audun2018subjective} based on the Dempster-Shafer framework~\cite{dempster1968generalization} for modeling of evidence. Different from the uncertainty of the estimated region mentioned above, some researchers~\cite{Ghesu2019, irvin2019chexpert, calli2019handling, pham2021interpreting} use different methods of training convolutional neural networks with uncertainty labels to approach experts' judgments on chest radiographs. Irvin et al.~\cite{irvin2019chexpert} first constructed a large chest radiograph dataset with uncertainty labels to automatically detect the presence of 14 observations in radiology reports. To address this issue, Pham et al.~\cite{pham2021interpreting} involved training cutting-edge convolutional neural networks that leverage hierarchical dependencies among abnormality labels. Additionally, they proposed incorporating the label smoothing technique to effectively handle uncertain samples, which constitute a substantial portion of nearly every Chest X-rays dataset. How to turn uncertain labels into definite labels to guide classification will be one of the hotspots of future research. After all, the cost of accurate labeling is too high and the time is longer.

In summary, although existing methods have made progress in evaluating prediction confidence for their respective tasks, there are still two limitations that require further improvement. Firstly, most previous studies on uncertainty in medical image classification utilize an MC-Dropout Bayesian-based approach, which are both stochastic and inefficient. Secondly, many of these studies are task-specific and lack end-to-end capability, making them less scalable. To overcome these limitations, it is worth exploring deterministic-based methods for medical image classification tasks, such as the evidential deep learning methods~\cite{evidential18, amini2020deep}. They employ the deterministic network to calculate final prediction and corresponding uncertainty score with a single forward pass without sampling. 

\begin{table*}[!t]
	\centering
	\caption{The summary of uncertainty estimation methods for medical image classification.}
 \begin{minipage}{\textwidth}
	\begin{tabular}{|p{40pt}|p{30pt}|p{50pt}|p{60pt}|p{160pt}|p{90pt}|}
		\hline
		Method            & Year   & Target       & Estimation & Dataset           & Use case         \\ \hline
  	\cite{Leibig2017} & 2017   & Fundus image & MC-Dropout & Messidor~\cite{decenciere2014feedback} & DR detection \\
        \cite{AYHAN2020101724}& 2020  & Fundus images & Bayesian-based & Kaggle DR\footnote{\url{https://www.kaggle.com/c/diabetic-retinopathy-detection}} \& IDRiD~\cite{porwal2018indian}  & DR detection \\
        \cite{ARAUJO2020101715}& 2020  & Fundus images  & Gaussian distribution centered & Kaggle DR~\footnote{\url{https://www.kaggle.com/c/diabetic-retinopathy-detection}}, Messidor-2~\cite{abramoff2016improved}, IDRID~\cite{porwal2018indian}, DMR~\cite{takahashi2017applying}, and SCREEN-DR (private dataset)  & DR detection \\
        \cite{Jaskari2022}& 2022  & Fundus images  & Bayesian-based  & EyePACS~\cite{cuadros2009eyepacs}, KSSHP~\footnote{\url{https://www.duodecimlehti.fi/duo15766}}, Messidor-2~\cite{abramoff2016improved}, and APTOS~\footnote{\url{https://www.kaggle.com/c/aptos2019-blindness-detection/overview/aptos-2019}}  & DR detection \\
        \cite{belharbi2021deep} & 2021 & Histology images & Bayesian-based & GlaS Dataset~\cite{sirinu2015stochastic} \& Camelyon16 Patch-Based Benchmark~\cite{rony2019deep} &  Breast cancer \\
        \cite{dolezal2022uncertainty} & 2022 & Histology images & Dropout & TCGA~\footnote{\url{https://portal.gdc.cancer.gov/projects/TCGA-LUSC}} \& CPTAC~\footnote{\url{https://www.cancerimagingarchive.net/collections/}} & Lung adenocarcinoma and lung squamous cell carcinoma\\
        \cite{feng2022trusted} & 2022 & Histology images & Evidential-based &  LKS dataset~\cite{maksoud2020sos} \& Fibroma (private dataset) &  Liver Kidney Stomach\\
        \cite{LINMANS2023102655}& 2023  & Histology images  & Deep ensemble & Camelyo17 challenge~\cite{bandi2018detection} and the PANDA challenge~\cite{bulten2022artificial} & OOD detection in digital pathology \\
        \cite{van2019quantifying}& 2019  & Dermoscopic image  & MC-Dropout & HAM10000 dataset~\cite{tschandl2018ham10000} & Skin lesion classification\\
        \cite{Combalia2020}& 2020  & Dermoscopic image  & MC-Dropout & ISIC2018 dataset~\cite{codella2019skin} and ISIC2019 dataset~\footnote{\url{https://www.kaggle.com/andrewm/isic-2019}}& Skin disease classification \\
        \cite{ABDAR2021104418}& 2021  & Dermoscopic image  & Deep ensemble &  Kaggle Skin Cancer dataset~\footnote{\url{https://www.kaggle.com/fanconic/skin-cancer-malignant-vs-benign}} \& ISIC2019 dataset~\footnote{\url{https://www.kaggle.com/andrewm/isic-2019}} & Skin disease classification \\ 
        \cite{Herzog2020}& 2020 & MRI & MC Dropout & private dataset & MRI based stroke analysis\\
        \cite{prince2023uncertainty}& 2023 & MRI & Variational inference & private dataset & MRI based Adamantinomatous Craniopharyngioma\\
	\cite{Ghesu2019,Ghesu2021} & 2019\&2021 & X-Ray & MC-Dropout & ChestX-Ray8~\cite{wang2017chestx} \& PLCO~\cite{gohagan2000prostate} &  Chest Radiograph Assessment \\
	\hline
	\end{tabular} 
	\label{tab_cls}%
\end{minipage}
\end{table*}%

\subsection{Segmentation}
Semantic segmentation, a crucial task in computer vision and image processing, involves assigning semantic labels to every pixel of an input image from a range of predefined classes~\cite{huang2019uncertainty}. There is a growing urgency in the context of semantic segmentation to explore ambiguity estimation in medical image pixels. Indeed, uncertainty in semantic segmentation can generally be divided into two types: (1) ambiguity within the area or boundary surrounding the tissue, and (2) unknown semantic categorization of the region or boundary. In medical domain, uncertainty estimation for medical image segmentation can be roughly divided into Bayesian-based ~\cite{2018probabilisticU,2019phiseg,2019supervisedUC,gantenbein2020revphiseg,wang2021medical,Sedai2018,carannante2021trustworthy} and Non-Bayesian-based methods~\cite{ruan2020mt,MIA2020exploringDropSeg,MIA2020polypssegUC,Yu2019,hu2022semi,wang2019aleatoric,krygier2021quantifying,2019ucBra,TMI20ensembleSeg,georgescu2022diversity,guo2022cardiac,huang2021belief,zou2022tbrats,huang2022evidence,judge2022reliable}. Bayesian-based methods enables the segmentation networks to learn a distribution over the network weights with uncertainty rather than a single pixel-wise estimate. To avoid computationally expensive by them, a variety of non-Bayesian methods have been developed, included Monte Carlo (MC) dropout based~\cite{ruan2020mt,MIA2020exploringDropSeg,MIA2020polypssegUC,Yu2019,hu2022semi,wang2019aleatoric,krygier2021quantifying}, ensemble-based~\cite{2019ucBra, TMI20ensembleSeg,georgescu2022diversity,guo2022cardiac} and  Determinnistic-based~\cite{huang2021belief,judge2022reliable,zou2022tbrats,huang2022evidence,Wang2023_TrFedDis,Wang2022_RJS}.

For the Bayesian-based methods, PU~\cite{2018probabilisticU} first considered the task of learning a distribution over segmentation given an input in medical domain. Other methods~\cite{2019phiseg,2019supervisedUC,gantenbein2020revphiseg} further improved the PU in terms of epistemic uncertainty and model efficiency. Sedai et al.~\cite{Sedai2018} used Bayesian deep learning for retinal layer segmentation with uncertainty quantification. Then, Carannante et al.~\cite{carannante2021trustworthy} used the first-order Taylor series approximation to propagate and learn the distribution of the model parameters for medical image segmentation.

To solve the problem of time overhead and accurate estimation of the posterior, the MC dropout-based methods were proposed. Nair et al.~\cite{MIA2020exploringDropSeg} first explored the multiple uncertainty estimates based on MC dropout in the context of deep networks for lesion detection and segmentation in medical image. Wickstr{\o}m et al.~\cite{MIA2020polypssegUC} developed MC dropout in FCN and model interpretability in the context of semantic segmentation of polyps from colonoscopy images. Yu et al.~\cite{Yu2019} introduced the MC dropout in a semi-supervised framework and presented an uncertainty-aware model for left atrium segmentation from 3D MR images. Wang et al.~\cite{wang2019aleatoric} developed test time augmentation method to analyze epistemic and aleatoric uncertainty for MC sampling-based medical image segmentation tasks at both pixel and structure levels. 

Another simple way to produce uncertainty for medical image segmentation is to use an ensemble of deep networks~\cite{2019ucBra, TMI20ensembleSeg,georgescu2022diversity,guo2022cardiac, cao2020uncertainty}. Mehrtash et al.~\cite{TMI20ensembleSeg} studied predictive uncertainty estimation by using multi-FCNs ensembling. Cao et al.~\cite{cao2020uncertainty} then developed an uncertainty aware model for semi-supervised ABUS mass segmentation based on ensemble learning. Guo et al.~\cite{guo2022cardiac} developed a globally optimal label fusion algorithm based on ensemble learning for short-axis cardiac MRI segmentation.

Unfortunately, the above methods cannot estimate the uncertainty of medical image segmentation with a single forward pass. Therefore, deterministic uncertainty estimation is proposed to train deterministic deep models with a single forward pass at test time. Amersfoort et al.~\cite{van2020uncertainty} exploited the ideas of radial basis function networks to devise deterministic uncertainty estimation. Mukhoti et al.~\cite{mukhoti2021deep} first extended deep deterministic uncertainty to semantic segmentation using feature space densities. Judge et al. ~\cite{judge2022reliable} implemented a contrastive method to learn a joint latent space which encodes a distribution of valid segmentations. Recently, evidential-based learning approaches~\cite{huang2021belief, zou2022tbrats, huang2022evidence} have been proposed for medical image segmentation due to their robustness and efficiency. As stated in~\cite{zou2022tbrats}, they treat neural network predictions as subjective opinions by parameterizing the class probabilities of the segmentation as a Dirichlet distribution. Huang et al.~\cite{huang2022evidence} computed a belief function at each voxel for each modality and then used Dempster’s rule for multi-modality medical image segmentation. 

In what follows, we briefly discuss the advantages and disadvantages of uncertainty estimation in medical image segmentation. Bayesian-based methods cleverly consider learning a distribution over segmentation given an input in medical domain, but their training process is complicated. The simple way to generate uncertainty for medical image segmentation is ensemble-based methods, but often require more training time and computational burden. MC dropout-based methods are the most common in medical image segmentation, but often requires multiple sampling to generate uncertainty through multiple forward passes. Recently, deterministic-based methods have attracted great attention, which only provide uncertainty for each pixel in medical images through the single forward pass. Although these methods sacrifice certain segmentation performance, their robustness and ability to detect OOD data are enhanced. In the future, there are many open research directions on uncertainty quantification in medical image segmentation that should be considered. First, how to generate more robust and calibrated uncertainty during the segmentation. Second, how to make better use of uncertainty to guide the improvement of segmentation performance. In addition, how to introduce uncertainty into the training process is also beneficial to the performance of medical image segmentation. In short, the application of uncertainty generation in medical image segmentation will be one of the emerging directions for reliable and explainable medical artificial intelligence. 

\begin{table*}[!t]
	\centering
	\caption{The summary of uncertainty estimation methods for medical image segmentation.}
	\begin{tabular}{|p{40pt}|p{30pt}|p{50pt}|p{60pt}|p{160pt}|p{90pt}|}
		\hline
		Method                     & Year & Target     & Estimation                   & Dataset              & Use case                                        \\ \hline
		\cite{Li2017}              & 2017 & 3D MRI     & MC-Dropout                   & ADNI dataset~\cite{saykin2015genetic}         & Volumetric segmentation                     \\
		\cite{DeVries2018}         & 2018 & skin image & MC-Dropout                   & ISIC 2017 dataset~\cite{isic2017}    & Skin lesion segmentation                    \\
		\cite{Eaton-Rosen2018}     & 2018 & 3D MRI     & MC-Dropout                   & BraTS 2017 dataset~\cite{menze2014multimodal} & Volumetric uncertainty                      \\
		\cite{Bragman2018}         & 2018 & 3D MR      & MC-Dropout                   & private dataset         & Multi-task for segmentation and regression  \\
		\cite{Garg2018}            & 2018 & 2D MRI     & MCMC sampling + Bayesian MRF & private dataset         & Brain image segmentation                    \\
		\cite{MIA2020exploringDropSeg}            & 2018 & 3D MRI     & MC-Dropout                   & private dataset         & Lesion segmentation, 4 uncertainty measures \\
		\cite{Jungo2018}           & 2018 & 2D MRI     & MC-Dropout                   & private dataset         & Brain Tumor Cavity Segmentation             \\
		\cite{Kwon2018}            & 2018 & 3D MRI     & MCMC sampling                & ISLES 2015 dataset~\cite{winzeck2018isles}   & Ischemic stroke lesion segmentation         \\
		\cite{Sedai2018}           & 2018 & OCT        & MC-Dropout                   & private dataset         & Retinal Layer Segmentation                  \\
		\cite{Jungo2018_MICCAI}    & 2018 & 2D MRI     & MC-Dropout                   & private dataset         & Inter-observer Variability                  \\
		\cite{Soberanis-Mukul2019} & 2019 & CT         & MC-Dropout                   & NIH pancreas dataset~\cite{roth2015deeporgan} & Organ Segmentation                          \\
		\cite{Seebock2019}         & 2019 & OCT        & MC-Dropout                   & private dataset         & Anomaly Detection                           \\
  		\cite{Yu2019} & 2019 & 3D MRI & MC-Dropout & Atrial Segmentation Challenge~\cite{Atrial} & Semi-supervised Segmentation \\
		\cite{wang2019aleatoric} & 2019 & MRI & Test-time augmentation &  private dataset   & Aleatoric uncertainty estimation \\
		\cite{2019phiseg} & 2019 & 2D CT, MRI & Probabilistic Unet & LIDC-IDRI lung CT~\cite{armato2011lung} and in-house prostate MR dataset & Multi-scale \\
		\cite{Awate2019}            & 2019 & 2D MRI     & MCMC sampling + Bayesian MRF & private dataset         & Brain image segmentation                    \\
  		\cite{ruan2020mt}            & 2020 & 2D CT     & MCMC sampling & private dataset         & renal tumor                   \\
            \cite{TMI20ensembleSeg}            & 2020 & 2D MRI \& 2D cine MR     & Ensemble & BraTS~\cite{menze2014multimodal}, ACDC~\cite{wolterink2017automatic}, PROSTATEx~\cite{litjens2014computer} and PROMISE12~\cite{litjens2014evaluation}   & Brain tumor \& Ventricular \& prostate  \\
            \cite{cao2020uncertainty} & 2020 & 2D ABUS \& 2D BUS   & Ensemble &  private dataset   & ABUS Mass Segmentation \\    \cite{huang2021belief} & 2021 & 3D MRI  & Evidential deep learning & BraTS 2018\&2019 dataset~\cite{menze2014multimodal}  & Belief function theory and evidential fusion  \\
	    \cite{zou2022tbrats}            & 2022 & 3D MRI     & Evidential deep learning & BraTS 2019 dataset~\cite{menze2014multimodal}  & Subjective logic theory and Dirichlet distribution \\
            \cite{judge2022reliable}  & 2022 & 2D US, X-ray  & Joint latent space & CAMUS~\cite{leclerc2019deep}, HMC-QU~\cite{degerli2021early}, Shenzen~\cite{jaeger2014two} and JSRT~\cite{shiraishi2000development} dataset & Cardiac and lung segmentation  \\

		 \hline
	\end{tabular} 
	\label{tab_seg}%
\end{table*}

\subsection{Other tasks}
As an important analytical tool for trustworthy learning, uncertainty estimation is also equipped in other medical image assessment tasks, such as image registration~\cite{Luo2019, LeFolgoc2017, Madsen2020, Luo2020_miccai, Markiewicz2021, Xu2022_miccai}, image reconstruction~\cite{Zhang2019_cvpr, Edupuganti2021, Sudarshan2021, Armanious2021, Narnhofer2022, Ezhov2022_miccai}, image denoising~\cite{Cui2022_miccai}, super-resolution~\cite{Tanno2017, Tanno2019, Tanno2021}, counting~\cite{eaton2019easy}, image detection~\cite{ Ozdemir2017} and tumor growth prediction~\cite{Petersen2019}.

Image registration is the foundation for many image-guided medical tasks. Estimating the uncertainty for image registration enables surgeons to assess the surgical risk based on the reliability of the registered image. If surgeons receive inaccurately calculated registration uncertainty and then misplace unwarranted confidence in the alignment results, severe consequences may result. Luo, et al~\cite{Luo2019} divided the registration uncertainty into two aspects: transformation uncertainty and label uncertainty. Le Folgoc, et al~\cite{LeFolgoc2017} investigated uncertainty quantification under a sparse Bayesian model of medical image registration. They implemented an exact inference scheme based on reversible jump Markov Chain Monte Carlo sampling to characterize the transformation posterior distribution. Madsen, et al~\cite{Madsen2020} and Luo et al.~\cite{Luo2020_miccai} also viewed surface registration as a probabilistic inference problem and use Gaussian Process Morphable Model as the prior model. Markiewicz et al.~\cite{Markiewicz2021} applied the uncertainty analysis to the multi-modal registration between PET and MRI images. Xu et al.~\cite{Xu2022_miccai} introduced the mean-teacher based registration framework. Instead of searching for a fixed weight, the teacher enables automatically adjusting the weights of the spatial regularization and temporal consistency regularization by taking advantage of the appearance uncertainty and the transformation uncertainty.

Image reconstruction is also the foundation task for medical image analysis. The goal of medical image reconstruction is to restore a high-fidelity image from partially observed measurements. Measuring the uncertainty in the process of reconstruction is critical. Zhang et al.~\cite{Zhang2019_cvpr} presented MRI reconstruction method that dynamically selects the measurements to take and iteratively refines the prediction in order to best reduce the reconstruction error and, thus, its uncertainty. Edupuganti et al.~\cite{Edupuganti2021} leveraged variational autoencoders to develop a probabilistic reconstruction scheme and exploit MC sampling to generate the uncertainty from the posterior of the image. Narnhofer et al.~\cite{Narnhofer2022} proposed a deterministic MRI Reconstruction which introduces a Bayesian framework for uncertainty quantification in single and multi-coil undersampled MRI reconstruction exploiting the total deep variation regularizer.

In addition, there are various efforts devoted to exploring uncertainty estimation for medical image analysis tasks. Tanno et al.~\cite{Tanno2017, Tanno2019, Tanno2021} focused on super-resolution and propose to account for intrinsic uncertainty through a heteroscedastic noise model and for parameter uncertainty through approximate Bayesian inference, and integrate them to quantify predictive uncertainty over the output. Cui et al.~\cite{Cui2022_miccai} introduced the uncertainty estimation into PET denoising task. They proposed a Nouveau variational autoencoder based model using quantile regression loss for simultaneous PET image denoising and uncertainty estimation. Eaton et al.~\cite{eaton2019easy} leveraged the counting task by introducing Predictive Intervals estimation  to calculate the counting intervals. Furthermore, Ozdemir et al.~\cite{Ozdemir2017} introduced uncertainty estimation into pulmonary nodule detection and Petersen et al.~\cite{Petersen2019} exploited it to the glioma growth. These attempts have demonstrated the importance of introducing uncertainty analysis into medical image analysis.

In general, as the fundamental tasks for medical image analysis, image registration and image reconstruction tasks have received the most attention when it comes to the application of uncertainty estimation. Estimating the uncertainty for these two tasks enables surgeons to assess the operative risk based on the trustworthiness of the registered or reconstructed image data. In addition, it is noted that uncertainty estimation has been less studied for other tasks such as image denoising, counting, and detection which are also important components of medical analysis. More future research work would be directed toward these tasks.




\begin{table*}[!t]
	\centering
	\caption{The summary of uncertainty estimation methods for the other medical image tasks.}
	\begin{tabular}{|p{40pt}|p{30pt}|p{50pt}|p{60pt}|p{160pt}|p{90pt}|}
		\hline
		Method             & Year   & Target & Estimation          & Dataset                                                     & Task             \\ \hline
  \cite{Luo2019}    & 2019 & 3D MRI & Entropy & CUMC12~\cite{CUMC12} and BraTS~\cite{menze2014multimodal} datasets & Image registration   \\
  \cite{Tanno2017,Tanno2019}   & 2017, 2019   & 3D MRI    & Variational Dropout & WU-Minn HCP~\cite{van2013wu}, Lifespan~\cite{lifespan}, Prisma~\cite{alexander2017image}, and Pathology~\cite{figini2018prediction} datasets & Super-resolution \\
		\cite{Ozdemir2017} & 2017   & 3D CT  & MC-Dropout & LUNA16~\cite{setio2017validation} dataset                                                    & Pulmonary Nodule Detection                \\
		\cite{eaton2019easy} & 2019 & 2D histopathological image & Bayesian-based: PI estimate & Cell histology~\cite{naylor2018segmentation} and WMH~\cite{kuijf2019standardized} datasets & Counting task \\
		\cite{Petersen2019} & 2019 & MRI & Probabilistic Unet & private dataset & Glioma Growth Prediction\\
		\cite{LeFolgoc2017}  & 2017  & 2D medical image  & MCMC & Private dataset  &  Image registration\\
   \cite{Zhang2019_cvpr}  & 2019  &  2D MRI & Bayesian-based & fastmri~\cite{zbontar2018fastmri} and ImageNet~\cite{deng2009imagenet} datasets  & Image reconstruction \\
      \cite{Madsen2020}  & 2020  &  CT & Gaussian process &  Public face~\cite{gerig2018morphable} and private femur bones datasets & Surface Registration \\
      \cite{Luo2020_miccai} & 2020  & MRI  & Gaussian process & RESECT~\cite{xiao2017resect} and private MIBS datasets  & Image Registration \\
\cite{Sudarshan2021} &  2021 & PET and MRI  & MC dropout & Private dataset  & Image reconstruction \\
\cite{Tanno2021} & 2021  &  3D MRI & Variational Dropout  & WU-Minn HCP~\cite{van2013wu}, Lifespan~\cite{lifespan}, Prisma~\cite{alexander2017image}, and Pathology~\cite{figini2018prediction} datasets  &  Super-resolution\\
\cite{Markiewicz2021}  & 2021  &  PET and MRI & MC sampling & Private dataset  &  Image registration\\
\cite{Edupuganti2021} &  2021 &  MRI &  MC sampling &  Mridata~\cite{mridata} & Image reconstruction \\
\cite{Armanious2021} & 2021  &  3D MRI & Gaussian process & OASIS-3 brain~\cite{lamontagne2019oasis} dataset & Biological age detection \\
\cite{Cui2022_miccai}  & 2022  & PET  & Quantile Regression & $^{11}$C-DASB\cite{ginovart200311c}  & PET Denoising \\
\cite{Xu2022_miccai} & 2022  & CT and MRI  & MC dropout & Private dataset  & Image registration \\
\cite{Narnhofer2022} & 2022  &  MRI & Deterministic & fastMRI~\cite{knoll2020fastmri} dataset  &  Image reconstruction\\
\cite{Ezhov2022_miccai} &  2022 & MRI and FET-PET  & MCMC &  BraTS~\cite{menze2014multimodal} and private dataset & Image reconstruction \\
		\hline
	\end{tabular} 
	\label{tab_recon}%
\end{table*}%




\section{Discussion and Conclusion}
 
Uncertainty estimation is a crucial aspect of deep learning in medical imaging, and it is an active area of research. However, there are several challenges and limitations associated with uncertainty estimation:
\begin{itemize}
\item \textbf{Lack of ground truth for uncertainty:} One of the main challenges in uncertainty estimation is the lack of ground truth for uncertainty in many medical applications. This makes it difficult to accurately estimate the uncertainty of deep learning models and evaluate the performance of different uncertainty estimation methods.
\item \textbf{Computational complexity:} The estimation of uncertainty in deep models can be computationally complex, especially for large and complex models. This can make it difficult to scale uncertainty estimation to use uncertainty estimation in real-time clinical systems.
\item \textbf{Trade-off between accuracy and reliability:} In practice, a model can have high accuracy but low reliability, or vice versa. Accuracy and reliability are complementary metrics that provide different perspectives on a model's performance, and both are important to consider when evaluating deep learning models. The goal should be to achieve high accuracy and reliability, but the trade-off between the two may vary depending on the specific use case and requirements of the model.
\item \textbf{Limited empirical evaluations}: Finally, there is a limited empirical evaluation of uncertainty estimation methods, especially in real clinical scenarios. This makes it difficult to compare and evaluate different methods, and to determine which methods are most effective and efficient in different tasks. However, one potential solution to address this issue is the utilization of different expert annotations, as demonstrated in~\cite{irvin2019chexpert}. It was annotated by different experts to capture the inherent uncertainty in the interpretation of radiographs. By leveraging such diverse expert annotations, it becomes possible to better understand and quantify uncertainty in clinical scenarios, facilitating more accurate evaluations of uncertainty estimation methods.
\end{itemize}   
Overall, these challenges and limitations need to be addressed in order to fully realize the potential of uncertainty estimation in deep learning.  By incorporating uncertainty estimation into AI systems, we can make them more robust and trustworthy in their predictions and decision-making processes, which can ultimately lead to improved patient outcomes.

\bibliographystyle{IEEEtran}
\bibliography{Ucerntainty_review}

\begin{thebibliography}{100}
\providecommand{\url}[1]{#1}
\csname url@samestyle\endcsname
\providecommand{\newblock}{\relax}
\providecommand{\bibinfo}[2]{#2}
\providecommand{\BIBentrySTDinterwordspacing}{\spaceskip=0pt\relax}
\providecommand{\BIBentryALTinterwordstretchfactor}{4}
\providecommand{\BIBentryALTinterwordspacing}{\spaceskip=\fontdimen2\font plus
\BIBentryALTinterwordstretchfactor\fontdimen3\font minus
  \fontdimen4\font\relax}
\providecommand{\BIBforeignlanguage}[2]{{%
\expandafter\ifx\csname l@#1\endcsname\relax
\typeout{** WARNING: IEEEtran.bst: No hyphenation pattern has been}%
\typeout{** loaded for the language `#1'. Using the pattern for}%
\typeout{** the default language instead.}%
\else
\language=\csname l@#1\endcsname
\fi
#2}}
\providecommand{\BIBdecl}{\relax}
\BIBdecl

\bibitem{Esteva2017}
A.~Esteva, B.~Kuprel, R.~A. Novoa, J.~Ko, S.~M. Swetter, H.~M. Blau, and
  S.~Thrun, ``{Dermatologist-level classification of skin cancer with deep
  neural networks},'' \emph{Nature}, vol. 542, no. 7639, pp. 115--118, feb
  2017.

\bibitem{Lee2019}
H.~Lee, S.~Yune, M.~Mansouri, M.~Kim, S.~H. Tajmir, C.~E. Guerrier, S.~A.
  Ebert, S.~R. Pomerantz, J.~M. Romero, S.~Kamalian, R.~G. Gonzalez, M.~H. Lev,
  and S.~Do, ``{An explainable deep-learning algorithm for the detection of
  acute intracranial haemorrhage from small datasets},'' \emph{Nature
  Biomedical Engineering}, vol.~3, no.~3, pp. 173--182, mar 2019.

\bibitem{Bi2019}
W.~L. Bi, A.~Hosny, M.~B. Schabath, M.~L. Giger, N.~J. Birkbak, A.~Mehrtash,
  T.~Allison, O.~Arnaout, C.~Abbosh, I.~F. Dunn, R.~H. Mak, R.~M. Tamimi, C.~M.
  Tempany, C.~Swanton, U.~Hoffmann, L.~H. Schwartz, R.~J. Gillies, R.~Y. Huang,
  and H.~J. W.~L. Aerts, ``{Artificial intelligence in cancer imaging: Clinical
  challenges and applications},'' \emph{CA: A Cancer Journal for Clinicians},
  vol.~69, no.~2, p. caac.21552, feb 2019.

\bibitem{Rajpurkar2022}
P.~Rajpurkar, E.~Chen, O.~Banerjee, and E.~J. Topol, ``{AI in health and
  medicine},'' \emph{Nature Medicine}, vol.~28, no.~1, pp. 31--38, jan 2022.

\bibitem{Shamshad2022}
F.~Shamshad, S.~Khan, S.~W. Zamir, M.~H. Khan, M.~Hayat, F.~S. Khan, and H.~Fu,
  ``{Transformers in Medical Imaging: A Survey},'' \emph{arXiv}, jan 2022.

\bibitem{fundus_survey}
T.~Li, W.~Bo, C.~Hu, H.~Kang, H.~Liu, K.~Wang, and H.~Fu, ``{Applications of
  deep learning in fundus images: A review},'' \emph{Medical Image Analysis},
  vol.~69, p. 101971, apr 2021.

\bibitem{Li2022}
B.~Li, P.~Qi, B.~Liu, S.~Di, J.~Liu, J.~Pei, J.~Yi, and B.~Zhou, ``{Trustworthy
  AI: From Principles to Practices},'' \emph{ACM Computing Surveys}, aug 2022.

\bibitem{Liang2022}
W.~Liang, G.~A. Tadesse, D.~Ho, L.~Fei-Fei, M.~Zaharia, C.~Zhang, and J.~Zou,
  ``{Advances, challenges and opportunities in creating data for trustworthy
  AI},'' \emph{Nature Machine Intelligence}, vol.~4, no.~8, pp. 669--677, aug
  2022.

\bibitem{Kendall2017}
A.~Kendall and Y.~Gal, ``{What Uncertainties Do We Need in Bayesian Deep
  Learning for Computer Vision?}'' in \emph{NIPS}, 2017.

\bibitem{Abdar2021}
M.~Abdar, F.~Pourpanah, S.~Hussain, D.~Rezazadegan, L.~Liu, M.~Ghavamzadeh,
  P.~Fieguth, X.~Cao, A.~Khosravi, U.~R. Acharya \emph{et~al.}, ``A review of
  uncertainty quantification in deep learning: Techniques, applications and
  challenges,'' \emph{Information Fusion}, vol.~76, pp. 243--297, 2021.

\bibitem{Li2022_CVPR}
B.~Li, Z.~Han, H.~Li, H.~Fu, and C.~Zhang, ``{Trustworthy Long-Tailed
  Classification},'' in \emph{CVPR}.\hskip 1em plus 0.5em minus 0.4em\relax
  IEEE, jun 2022, pp. 6960--6969.

\bibitem{der2009aleatory}
A.~Der~Kiureghian and O.~Ditlevsen, ``Aleatory or epistemic? does it matter?''
  \emph{Structural safety}, vol.~31, no.~2, pp. 105--112, 2009.

\bibitem{evidential18}
M.~Sensoy, L.~Kaplan, and M.~Kandemir, ``Evidential deep learning to quantify
  classification uncertainty,'' in \emph{Proceedings of the 32nd International
  Conference on Neural Information Processing Systems}, 2018, pp. 3183--3193.

\bibitem{9767662}
Z.~Han, C.~Zhang, H.~Fu, and J.~T. Zhou, ``Trusted multi-view classification
  with dynamic evidential fusion,'' \emph{IEEE Transactions on Pattern Analysis
  and Machine Intelligence}, vol.~45, no.~2, pp. 2551--2566, 2023.

\bibitem{Han2021_ICLR}
------, ``{Trusted Multi-View Classification},'' in \emph{ICLR}, 2021.

\bibitem{Ma2021_NIPS}
H.~Ma, C.~Zhang, J.~Zhou, Z.~Han, H.~Fu, and Q.~Hu, ``{Trustworthy multimodal
  regression with mixture of normal-inverse gamma distributions},'' in
  \emph{NeurIPS}, 2021.

\bibitem{Zou2023}
K.~Zou, X.~Yuan, X.~Shen, Y.~Chen, M.~Wang, R.~S.~M. Goh, Y.~Liu, and H.~Fu,
  ``{EvidenceCap: Towards trustworthy medical image segmentation via evidential
  identity cap},'' \emph{arXiv}, jan 2023.

\bibitem{2020NIPSdeterministic}
J.~Liu, Z.~Lin, S.~Padhy, D.~Tran, T.~Bedrax~Weiss, and B.~Lakshminarayanan,
  ``Simple and principled uncertainty estimation with deterministic deep
  learning via distance awareness,'' \emph{Advances in Neural Information
  Processing Systems}, vol.~33, pp. 7498--7512, 2020.

\bibitem{mukhoti2021deep}
J.~Mukhoti, J.~van Amersfoort, P.~H. Torr, and Y.~Gal, ``Deep deterministic
  uncertainty for semantic segmentation,'' \emph{arXiv preprint
  arXiv:2111.00079}, 2021.

\bibitem{van2020uncertainty}
J.~Van~Amersfoort, L.~Smith, Y.~W. Teh, and Y.~Gal, ``Uncertainty estimation
  using a single deep deterministic neural network,'' in \emph{International
  conference on machine learning}.\hskip 1em plus 0.5em minus 0.4em\relax PMLR,
  2020, pp. 9690--9700.

\bibitem{van2017neural}
A.~Van Den~Oord, O.~Vinyals \emph{et~al.}, ``Neural discrete representation
  learning,'' \emph{Advances in neural information processing systems},
  vol.~30, 2017.

\bibitem{razavi2019generating}
A.~Razavi, A.~Van~den Oord, and O.~Vinyals, ``Generating diverse high-fidelity
  images with vq-vae-2,'' \emph{Advances in neural information processing
  systems}, vol.~32, 2019.

\bibitem{maddox2019simple}
W.~J. Maddox, P.~Izmailov, T.~Garipov, D.~P. Vetrov, and A.~G. Wilson, ``A
  simple baseline for bayesian uncertainty in deep learning,'' \emph{Advances
  in Neural Information Processing Systems}, vol.~32, 2019.

\bibitem{dera2021premium}
D.~Dera, N.~C. Bouaynaya, G.~Rasool, R.~Shterenberg, and H.~M.
  Fathallah-Shaykh, ``Premium-cnn: Propagating uncertainty towards robust
  convolutional neural networks,'' \emph{IEEE Transactions on Signal
  Processing}, vol.~69, pp. 4669--4684, 2021.

\bibitem{srivastava2014dropout}
N.~Srivastava, G.~Hinton, A.~Krizhevsky, I.~Sutskever, and R.~Salakhutdinov,
  ``Dropout: a simple way to prevent neural networks from overfitting,''
  \emph{The journal of machine learning research}, vol.~15, no.~1, pp.
  1929--1958, 2014.

\bibitem{ensemble17}
B.~Lakshminarayanan, A.~Pritzel, and C.~Blundell, ``Simple and scalable
  predictive uncertainty estimation using deep ensembles,'' \emph{Advances in
  Neural Information Processing Systems}, vol.~30, 2017.

\bibitem{2019EnsembleTTA}
A.~Ashukha, A.~Lyzhov, D.~Molchanov, and D.~Vetrov, ``Pitfalls of in-domain
  uncertainty estimation and ensembling in deep learning,'' in
  \emph{International Conference on Learning Representations}, 2019.

\bibitem{ashukha2020pitfalls}
------, ``Pitfalls of in-domain uncertainty estimation and ensembling in deep
  learning,'' in \emph{International Conference on Learning Representations},
  2020.

\bibitem{shen2022DCA}
Y.~Shen and D.~Cremers, ``Deep combinatorial aggregation,'' \emph{arXiv
  preprint arXiv:2210.06436}, 2022.

\bibitem{Ezhov2022_miccai}
I.~Ezhov, T.~Mot, S.~Shit, J.~Lipkova, J.~C. Paetzold, F.~Kofler,
  C.~Pellegrini, M.~Kollovieh, F.~Navarro, H.~Li, M.~Metz, B.~Wiestler, and
  B.~Menze, ``{Geometry-Aware Neural Solver for Fast Bayesian Calibration of
  Brain Tumor Models},'' \emph{IEEE Transactions on Medical Imaging}, vol.~41,
  no.~5, pp. 1269--1278, may 2022.

\bibitem{WANG201934}
G.~Wang, W.~Li, M.~Aertsen, J.~Deprest, S.~Ourselin, and T.~Vercauteren,
  ``Aleatoric uncertainty estimation with test-time augmentation for medical
  image segmentation with convolutional neural networks,''
  \emph{Neurocomputing}, vol. 338, pp. 34--45, 2019.

\bibitem{9413010}
K.~Patel, W.~Beluch, D.~Zhang, M.~Pfeiffer, and B.~Yang, ``On-manifold
  adversarial data augmentation improves uncertainty calibration,'' in
  \emph{International Conference on Pattern Recognition (ICPR)}, 2021.

\bibitem{Leibig2017}
C.~Leibig, V.~Allken, M.~S. Ayhan, P.~Berens, and S.~Wahl, ``{Leveraging
  uncertainty information from deep neural networks for disease detection},''
  \emph{Scientific Reports}, vol.~7, no.~1, p. 17816, 2017.

\bibitem{ARAUJO2020101715}
T.~Araújo, G.~Aresta, L.~Mendonça, S.~Penas, C.~Maia, Ângela Carneiro, A.~M.
  Mendonça, and A.~Campilho, ``Dr|graduate: Uncertainty-aware deep
  learning-based diabetic retinopathy grading in eye fundus images,''
  \emph{Medical Image Analysis}, vol.~63, p. 101715, 2020.

\bibitem{AYHAN2020101724}
M.~S. Ayhan, L.~Kühlewein, G.~Aliyeva, W.~Inhoffen, F.~Ziemssen, and
  P.~Berens, ``Expert-validated estimation of diagnostic uncertainty for deep
  neural networks in diabetic retinopathy detection,'' \emph{Medical Image
  Analysis}, vol.~64, p. 101724, 2020.

\bibitem{Jaskari2022}
J.~Jaskari, J.~Sahlsten, T.~Damoulas, J.~Knoblauch, S.~Särkkä,
  L.~Kärkkäinen, K.~Hietala, and K.~K. Kaski, ``Uncertainty-aware deep
  learning methods for robust diabetic retinopathy classification,'' \emph{IEEE
  Access}, vol.~10, pp. 76\,669--76\,681, 2022.

\bibitem{belharbi2021deep}
S.~Belharbi, J.~Rony, J.~Dolz, I.~B. Ayed, L.~McCaffrey, and E.~Granger, ``Deep
  interpretable classification and weakly-supervised segmentation of histology
  images via max-min uncertainty,'' \emph{IEEE Transactions on Medical
  Imaging}, vol.~41, no.~3, pp. 702--714, 2021.

\bibitem{dolezal2022uncertainty}
J.~M. Dolezal, A.~Srisuwananukorn, D.~Karpeyev, S.~Ramesh, S.~Kochanny,
  B.~Cody, A.~S. Mansfield, S.~Rakshit, R.~Bansal, M.~C. Bois \emph{et~al.},
  ``Uncertainty-informed deep learning models enable high-confidence
  predictions for digital histopathology,'' \emph{Nature communications},
  vol.~13, no.~1, p. 6572, 2022.

\bibitem{feng2022trusted}
M.~Feng, K.~Xu, N.~Wu, W.~Huang, Y.~Bai, C.~Wang, and H.~Wang, ``Trusted
  multi-scale classification framework for whole slide image,'' \emph{arXiv
  preprint arXiv:2207.05290}, 2022.

\bibitem{LINMANS2023102655}
J.~Linmans, S.~Elfwing, J.~{van der Laak}, and G.~Litjens, ``Predictive
  uncertainty estimation for out-of-distribution detection in digital
  pathology,'' \emph{Medical Image Analysis}, vol.~83, p. 102655, 2023.

\bibitem{van2019quantifying}
P.~Van~Molle, T.~Verbelen, C.~De~Boom, B.~Vankeirsbilck, J.~De~Vylder,
  B.~Diricx, T.~Kimpe, P.~Simoens, and B.~Dhoedt, ``Quantifying uncertainty of
  deep neural networks in skin lesion classification,'' in \emph{Uncertainty
  for Safe Utilization of Machine Learning in Medical Imaging and Clinical
  Image-Based Procedures}.\hskip 1em plus 0.5em minus 0.4em\relax Springer,
  2019, pp. 52--61.

\bibitem{Combalia2020}
M.~Combalia, F.~Hueto, S.~Puig, J.~Malvehy, and V.~Vilaplana, ``Uncertainty
  estimation in deep neural networks for dermoscopic image classification,'' in
  \emph{Proceedings of the IEEE/CVF Conference on Computer Vision and Pattern
  Recognition (CVPR) Workshops}, June 2020.

\bibitem{ABDAR2021104418}
M.~Abdar, M.~Samami, S.~{Dehghani Mahmoodabad}, T.~Doan, B.~Mazoure,
  R.~Hashemifesharaki, L.~Liu, A.~Khosravi, U.~R. Acharya, V.~Makarenkov, and
  S.~Nahavandi, ``Uncertainty quantification in skin cancer classification
  using three-way decision-based bayesian deep learning,'' \emph{Computers in
  Biology and Medicine}, vol. 135, p. 104418, 2021.

\bibitem{Herzog2020}
L.~Herzog, E.~Murina, O.~D{\"{u}}rr, S.~Wegener, and B.~Sick, ``{Integrating
  uncertainty in deep neural networks for MRI based stroke analysis},''
  \emph{Medical Image Analysis}, vol.~65, p. 101790, oct 2020.

\bibitem{prince2023uncertainty}
E.~W. Prince, D.~Ghosh, C.~G{\"o}rg, and T.~C. Hankinson, ``Uncertainty-aware
  deep learning classification of adamantinomatous craniopharyngioma from
  preoperative mri,'' \emph{Diagnostics}, vol.~13, no.~6, p. 1132, 2023.

\bibitem{Ghesu2019}
F.~C. Ghesu, B.~Georgescu, E.~Gibson, S.~Guendel, M.~K. Kalra, R.~Singh, S.~R.
  Digumarthy, S.~Grbic, and D.~Comaniciu, ``Quantifying and leveraging
  classification uncertainty for chest radiograph assessment,'' in
  \emph{International Conference on Medical Image Computing and
  Computer-Assisted Intervention}.\hskip 1em plus 0.5em minus 0.4em\relax
  Springer, 2019, pp. 676--684.

\bibitem{irvin2019chexpert}
J.~Irvin, P.~Rajpurkar, M.~Ko, Y.~Yu, S.~Ciurea-Ilcus, C.~Chute, H.~Marklund,
  B.~Haghgoo, R.~Ball, K.~Shpanskaya \emph{et~al.}, ``Chexpert: A large chest
  radiograph dataset with uncertainty labels and expert comparison,'' in
  \emph{Proceedings of the AAAI conference on artificial intelligence},
  vol.~33, no.~01, 2019, pp. 590--597.

\bibitem{calli2019handling}
E.~Calli, E.~Sogancioglu, E.~T. Scholten, K.~Murphy, and B.~van Ginneken,
  ``Handling label noise through model confidence and uncertainty: application
  to chest radiograph classification,'' in \emph{Medical Imaging 2019:
  Computer-Aided Diagnosis}, vol. 10950.\hskip 1em plus 0.5em minus 0.4em\relax
  SPIE, 2019, pp. 289--296.

\bibitem{pham2021interpreting}
H.~H. Pham, T.~T. Le, D.~Q. Tran, D.~T. Ngo, and H.~Q. Nguyen, ``Interpreting
  chest x-rays via cnns that exploit hierarchical disease dependencies and
  uncertainty labels,'' \emph{Neurocomputing}, vol. 437, pp. 186--194, 2021.

\bibitem{maksoud2020sos}
S.~Maksoud, K.~Zhao, P.~Hobson, A.~Jennings, and B.~C. Lovell, ``Sos: Selective
  objective switch for rapid immunofluorescence whole slide image
  classification,'' in \emph{Proceedings of the IEEE/CVF Conference on Computer
  Vision and Pattern Recognition}, 2020, pp. 3862--3871.

\bibitem{audun2018subjective}
A.~JSANG, \emph{Subjective Logic: A formalism for reasoning under
  uncertainty}.\hskip 1em plus 0.5em minus 0.4em\relax Springer, 2018.

\bibitem{dempster1968generalization}
A.~P. Dempster, ``A generalization of bayesian inference,'' \emph{Journal of
  the Royal Statistical Society: Series B (Methodological)}, vol.~30, no.~2,
  pp. 205--232, 1968.

\bibitem{amini2020deep}
A.~Amini, W.~Schwarting, A.~Soleimany, and D.~Rus, ``Deep evidential
  regression,'' \emph{Advances in Neural Information Processing Systems},
  vol.~33, pp. 14\,927--14\,937, 2020.

\bibitem{decenciere2014feedback}
E.~Decenci{\`e}re, X.~Zhang, G.~Cazuguel, B.~Lay, B.~Cochener, C.~Trone,
  P.~Gain, R.~Ordonez, P.~Massin, A.~Erginay \emph{et~al.}, ``Feedback on a
  publicly distributed image database: the messidor database,'' \emph{Image
  Analysis \& Stereology}, vol.~33, no.~3, pp. 231--234, 2014.

\bibitem{porwal2018indian}
P.~Porwal, S.~Pachade, R.~Kamble, M.~Kokare, G.~Deshmukh, V.~Sahasrabuddhe, and
  F.~Meriaudeau, ``Indian diabetic retinopathy image dataset (idrid): a
  database for diabetic retinopathy screening research,'' \emph{Data}, vol.~3,
  no.~3, p.~25, 2018.

\bibitem{abramoff2016improved}
M.~D. Abr{\`a}moff, Y.~Lou, A.~Erginay, W.~Clarida, R.~Amelon, J.~C. Folk, and
  M.~Niemeijer, ``Improved automated detection of diabetic retinopathy on a
  publicly available dataset through integration of deep learning,''
  \emph{Investigative ophthalmology \& visual science}, vol.~57, no.~13, pp.
  5200--5206, 2016.

\bibitem{takahashi2017applying}
H.~Takahashi, H.~Tampo, Y.~Arai, Y.~Inoue, and H.~Kawashima, ``Applying
  artificial intelligence to disease staging: Deep learning for improved
  staging of diabetic retinopathy,'' \emph{PloS one}, vol.~12, no.~6, p.
  e0179790, 2017.

\bibitem{cuadros2009eyepacs}
J.~Cuadros and G.~Bresnick, ``Eyepacs: an adaptable telemedicine system for
  diabetic retinopathy screening,'' \emph{Journal of diabetes science and
  technology}, vol.~3, no.~3, pp. 509--516, 2009.

\bibitem{sirinu2015stochastic}
K.~Sirinukunwattana, D.~R. Snead, and N.~M. Rajpoot, ``A stochastic polygons
  model for glandular structures in colon histology images,'' \emph{IEEE
  transactions on medical imaging}, vol.~34, no.~11, pp. 2366--2378, 2015.

\bibitem{rony2019deep}
J.~Rony, S.~Belharbi, J.~Dolz, I.~B. Ayed, L.~McCaffrey, and E.~Granger, ``Deep
  weakly-supervised learning methods for classification and localization in
  histology images: a survey,'' \emph{arXiv preprint arXiv:1909.03354}, 2019.

\bibitem{bandi2018detection}
P.~Bandi, O.~Geessink, Q.~Manson, M.~Van~Dijk, M.~Balkenhol, M.~Hermsen, B.~E.
  Bejnordi, B.~Lee, K.~Paeng, A.~Zhong \emph{et~al.}, ``From detection of
  individual metastases to classification of lymph node status at the patient
  level: the camelyon17 challenge,'' \emph{IEEE transactions on medical
  imaging}, vol.~38, no.~2, pp. 550--560, 2018.

\bibitem{bulten2022artificial}
W.~Bulten, K.~Kartasalo, P.-H.~C. Chen, P.~Str{\"o}m, H.~Pinckaers, K.~Nagpal,
  Y.~Cai, D.~F. Steiner, H.~van Boven, R.~Vink \emph{et~al.}, ``Artificial
  intelligence for diagnosis and gleason grading of prostate cancer: the panda
  challenge,'' \emph{Nature medicine}, vol.~28, no.~1, pp. 154--163, 2022.

\bibitem{tschandl2018ham10000}
P.~Tschandl, C.~Rosendahl, and H.~Kittler, ``The ham10000 dataset, a large
  collection of multi-source dermatoscopic images of common pigmented skin
  lesions,'' \emph{Scientific data}, vol.~5, no.~1, pp. 1--9, 2018.

\bibitem{codella2019skin}
N.~Codella, V.~Rotemberg, P.~Tschandl, M.~E. Celebi, S.~Dusza, D.~Gutman,
  B.~Helba, A.~Kalloo, K.~Liopyris, M.~Marchetti \emph{et~al.}, ``Skin lesion
  analysis toward melanoma detection 2018: A challenge hosted by the
  international skin imaging collaboration (isic),'' \emph{arXiv preprint
  arXiv:1902.03368}, 2019.

\bibitem{Ghesu2021}
F.~C. Ghesu, B.~Georgescu, A.~Mansoor, Y.~Yoo, E.~Gibson, R.~Vishwanath,
  A.~Balachandran, J.~M. Balter, Y.~Cao, R.~Singh, S.~R. Digumarthy, M.~K.
  Kalra, S.~Grbic, and D.~Comaniciu, ``{Quantifying and leveraging predictive
  uncertainty for medical image assessment},'' \emph{Medical Image Analysis},
  vol.~68, p. 101855, feb 2021.

\bibitem{wang2017chestx}
X.~Wang, Y.~Peng, L.~Lu, Z.~Lu, M.~Bagheri, and R.~M. Summers, ``Chestx-ray8:
  Hospital-scale chest x-ray database and benchmarks on weakly-supervised
  classification and localization of common thorax diseases,'' in
  \emph{Proceedings of the IEEE conference on computer vision and pattern
  recognition}, 2017, pp. 2097--2106.

\bibitem{gohagan2000prostate}
J.~K. Gohagan, P.~C. Prorok, R.~B. Hayes, B.-S. Kramer, P.~P. Team
  \emph{et~al.}, ``The prostate, lung, colorectal and ovarian (plco) cancer
  screening trial of the national cancer institute: history, organization, and
  status,'' \emph{Controlled clinical trials}, vol.~21, no.~6, pp. 251S--272S,
  2000.

\bibitem{huang2019uncertainty}
Y.-H. Huang, M.~Proesmans, S.~Georgoulis, and L.~Van~Gool, ``Uncertainty based
  model selection for fast semantic segmentation,'' in \emph{2019 16th
  International Conference on Machine Vision Applications (MVA)}.\hskip 1em
  plus 0.5em minus 0.4em\relax IEEE, 2019, pp. 1--6.

\bibitem{2018probabilisticU}
S.~Kohl, B.~Romera-Paredes, C.~Meyer, J.~De~Fauw, J.~R. Ledsam, K.~Maier-Hein,
  S.~Eslami, D.~Jimenez~Rezende, and O.~Ronneberger, ``A probabilistic u-net
  for segmentation of ambiguous images,'' \emph{Advances in Neural Information
  Processing Systems}, vol.~31, 2018.

\bibitem{2019phiseg}
C.~F. Baumgartner, K.~C. Tezcan, K.~Chaitanya, A.~M. H{\"o}tker, U.~J.
  Muehlematter, K.~Schawkat, A.~S. Becker, O.~Donati, and E.~Konukoglu,
  ``Phiseg: Capturing uncertainty in medical image segmentation,'' in
  \emph{International Conference on Medical Image Computing and
  Computer-Assisted Intervention}.\hskip 1em plus 0.5em minus 0.4em\relax
  Springer, 2019, pp. 119--127.

\bibitem{2019supervisedUC}
S.~Hu, D.~Worrall, S.~Knegt, B.~Veeling, H.~Huisman, and M.~Welling,
  ``Supervised uncertainty quantification for segmentation with multiple
  annotations,'' in \emph{International Conference on Medical Image Computing
  and Computer-Assisted Intervention}.\hskip 1em plus 0.5em minus 0.4em\relax
  Springer, 2019, pp. 137--145.

\bibitem{gantenbein2020revphiseg}
M.~Gantenbein, E.~Erdil, and E.~Konukoglu, ``Revphiseg: A memory-efficient
  neural network for uncertainty quantification in medical image
  segmentation,'' in \emph{Uncertainty for Safe Utilization of Machine Learning
  in Medical Imaging, and Graphs in Biomedical Image Analysis}.\hskip 1em plus
  0.5em minus 0.4em\relax Springer, 2020, pp. 13--22.

\bibitem{wang2021medical}
L.~Wang, L.~Ju, D.~Zhang, X.~Wang, W.~He, Y.~Huang, Z.~Yang, X.~Yao, X.~Zhao,
  X.~Ye \emph{et~al.}, ``Medical matting: a new perspective on medical
  segmentation with uncertainty,'' in \emph{International Conference on Medical
  Image Computing and Computer-Assisted Intervention}.\hskip 1em plus 0.5em
  minus 0.4em\relax Springer, 2021, pp. 573--583.

\bibitem{Sedai2018}
S.~Sedai, B.~Antony, D.~Mahapatra, and R.~Garnavi, ``Joint segmentation and
  uncertainty visualization of retinal layers in optical coherence tomography
  images using bayesian deep learning,'' in \emph{Computational Pathology and
  Ophthalmic Medical Image Analysis}.\hskip 1em plus 0.5em minus 0.4em\relax
  Springer, 2018, pp. 219--227.

\bibitem{carannante2021trustworthy}
G.~Carannante, D.~Dera, N.~C. Bouaynaya, R.~Ghulam, and H.~M. Fathallah-Shaykh,
  ``Trustworthy medical segmentation with uncertainty estimation,'' \emph{arXiv
  preprint arXiv:2111.05978}, 2021.

\bibitem{ruan2020mt}
Y.~Ruan, D.~Li, H.~Marshall, T.~Miao, T.~Cossetto, I.~Chan, O.~Daher,
  F.~Accorsi, A.~Goela, and S.~Li, ``Mt-ucgan: Multi-task
  uncertainty-constrained gan for joint segmentation, quantification and
  uncertainty estimation of renal tumors on ct,'' in \emph{International
  Conference on Medical Image Computing and Computer-Assisted
  Intervention}.\hskip 1em plus 0.5em minus 0.4em\relax Springer, 2020, pp.
  439--449.

\bibitem{MIA2020exploringDropSeg}
T.~Nair, d.~Precup, D.~L. Arnold, and T.~Arbel, ``Exploring uncertainty
  measures in deep networks for multiple sclerosis lesion detection and
  segmentation,'' \emph{Medical image analysis}, vol.~59, p. 101557, 2020.

\bibitem{MIA2020polypssegUC}
K.~Wickstr{\o}m, M.~Kampffmeyer, and R.~Jenssen, ``Uncertainty and
  interpretability in convolutional neural networks for semantic segmentation
  of colorectal polyps,'' \emph{Medical image analysis}, vol.~60, p. 101619,
  2020.

\bibitem{Yu2019}
L.~Yu, S.~Wang, X.~Li, C.-W. Fu, and P.-A. Heng, ``Uncertainty-aware
  self-ensembling model for semi-supervised 3d left atrium segmentation,'' in
  \emph{International Conference on Medical Image Computing and
  Computer-Assisted Intervention}.\hskip 1em plus 0.5em minus 0.4em\relax
  Springer, 2019, pp. 605--613.

\bibitem{hu2022semi}
L.~Hu, J.~Li, X.~Peng, J.~Xiao, B.~Zhan, C.~Zu, X.~Wu, J.~Zhou, and Y.~Wang,
  ``Semi-supervised npc segmentation with uncertainty and attention guided
  consistency,'' \emph{Knowledge-Based Systems}, vol. 239, p. 108021, 2022.

\bibitem{wang2019aleatoric}
G.~Wang, W.~Li, M.~Aertsen, J.~Deprest, S.~Ourselin, and T.~Vercauteren,
  ``Aleatoric uncertainty estimation with test-time augmentation for medical
  image segmentation with convolutional neural networks,''
  \emph{Neurocomputing}, vol. 338, pp. 34--45, 2019.

\bibitem{krygier2021quantifying}
M.~C. Krygier, T.~LaBonte, C.~Martinez, C.~Norris, K.~Sharma, L.~N. Collins,
  P.~P. Mukherjee, and S.~A. Roberts, ``Quantifying the unknown impact of
  segmentation uncertainty on image-based simulations,'' \emph{Nature
  communications}, vol.~12, no.~1, pp. 1--11, 2021.

\bibitem{2019ucBra}
R.~McKinley, M.~Rebsamen, R.~Meier, and R.~Wiest, ``Triplanar ensemble of
  3d-to-2d cnns with label-uncertainty for brain tumor segmentation,'' in
  \emph{International MICCAI Brainlesion Workshop}.\hskip 1em plus 0.5em minus
  0.4em\relax Springer, 2019, pp. 379--387.

\bibitem{TMI20ensembleSeg}
A.~Mehrtash, W.~M. Wells, C.~M. Tempany, P.~Abolmaesumi, and T.~Kapur,
  ``Confidence calibration and predictive uncertainty estimation for deep
  medical image segmentation,'' \emph{IEEE Transactions on Medical Imaging},
  vol.~39, no.~12, pp. 3868--3878, 2020.

\bibitem{georgescu2022diversity}
M.-I. Georgescu, R.~T. Ionescu, and A.-I. Miron, ``Diversity-promoting ensemble
  for medical image segmentation,'' \emph{arXiv preprint arXiv:2210.12388},
  2022.

\bibitem{guo2022cardiac}
F.~Guo, M.~Ng, G.~Kuling, and G.~Wright, ``Cardiac mri segmentation with sparse
  annotations: Ensembling deep learning uncertainty and shape priors,''
  \emph{Medical Image Analysis}, vol.~81, p. 102532, 2022.

\bibitem{huang2021belief}
L.~Huang, S.~Ruan, and T.~Denoeux, ``Belief function-based semi-supervised
  learning for brain tumor segmentation,'' in \emph{2021 IEEE 18th
  International Symposium on Biomedical Imaging (ISBI)}.\hskip 1em plus 0.5em
  minus 0.4em\relax IEEE, 2021, pp. 160--164.

\bibitem{zou2022tbrats}
K.~Zou, X.~Yuan, X.~Shen, M.~Wang, and H.~Fu, ``Tbrats: Trusted brain tumor
  segmentation,'' in \emph{International Conference on Medical Image Computing
  and Computer-Assisted Intervention}.\hskip 1em plus 0.5em minus 0.4em\relax
  Springer, 2022, pp. 503--513.

\bibitem{huang2022evidence}
L.~Huang, T.~Denoeux, P.~Vera, and S.~Ruan, ``Evidence fusion with contextual
  discounting for multi-modality medical image segmentation,'' in
  \emph{International Conference on Medical Image Computing and
  Computer-Assisted Intervention}.\hskip 1em plus 0.5em minus 0.4em\relax
  Springer, 2022, pp. 401--411.

\bibitem{judge2022reliable}
T.~Judge, O.~Bernard, M.~Porumb, A.~Chartsias, A.~Beqiri, and P.-M. Jodoin,
  ``Crisp - reliable uncertainty estimation for medical image segmentation,''
  in \emph{International Conference on Medical Image Computing and
  Computer-Assisted Intervention}.\hskip 1em plus 0.5em minus 0.4em\relax
  Springer, 2022, pp. 492--502.

\bibitem{Wang2023_TrFedDis}
M.~Wang, K.~Yu, C.-M. Feng, Y.~Qian, K.~Zou, L.~Wang, R.~S.~M. Goh, X.~Xu,
  Y.~Liu, and H.~Fu, ``{TrFedDis: Trusted Federated Disentangling Network for
  Non-IID Domain Feature},'' \emph{arXiv}, jan 2023.

\bibitem{Wang2022_RJS}
M.~Wang, K.~Yu, C.-M. Feng, K.~Zou, Y.~Xu, Q.~Meng, R.~S.~M. Goh, Y.~Liu,
  X.~Xu, and H.~Fu, ``{Reliable Joint Segmentation of Retinal Edema Lesions in
  OCT Images},'' \emph{arXiv}, dec 2022.

\bibitem{cao2020uncertainty}
X.~Cao, H.~Chen, Y.~Li, Y.~Peng, S.~Wang, and L.~Cheng, ``Uncertainty aware
  temporal-ensembling model for semi-supervised abus mass segmentation,''
  \emph{IEEE transactions on medical imaging}, vol.~40, no.~1, pp. 431--443,
  2020.

\bibitem{Li2017}
W.~Li, G.~Wang, L.~Fidon, S.~Ourselin, M.~J. Cardoso, and T.~Vercauteren, ``On
  the compactness, efficiency, and representation of 3d convolutional networks:
  brain parcellation as a pretext task,'' in \emph{International conference on
  information processing in medical imaging}.\hskip 1em plus 0.5em minus
  0.4em\relax Springer, 2017, pp. 348--360.

\bibitem{saykin2015genetic}
A.~J. Saykin, L.~Shen, X.~Yao, S.~Kim, K.~Nho, S.~L. Risacher, V.~K. Ramanan,
  T.~M. Foroud, K.~M. Faber, N.~Sarwar \emph{et~al.}, ``Genetic studies of
  quantitative mci and ad phenotypes in adni: progress, opportunities, and
  plans,'' \emph{Alzheimer's \& Dementia}, vol.~11, no.~7, pp. 792--814, 2015.

\bibitem{DeVries2018}
T.~DeVries and G.~W. Taylor, ``Leveraging uncertainty estimates for predicting
  segmentation quality,'' \emph{arXiv preprint arXiv:1807.00502}, 2018.

\bibitem{isic2017}
N.~C. Codella, D.~Gutman, M.~E. Celebi, B.~Helba, M.~A. Marchetti, S.~W. Dusza,
  A.~Kalloo, K.~Liopyris, N.~Mishra, H.~Kittler \emph{et~al.}, ``Skin lesion
  analysis toward melanoma detection: A challenge at the 2017 international
  symposium on biomedical imaging (isbi), hosted by the international skin
  imaging collaboration (isic),'' in \emph{2018 IEEE 15th international
  symposium on biomedical imaging (ISBI 2018)}.\hskip 1em plus 0.5em minus
  0.4em\relax IEEE, 2018, pp. 168--172.

\bibitem{Eaton-Rosen2018}
Z.~Eaton-Rosen, F.~Bragman, S.~Bisdas, S.~Ourselin, and M.~J. Cardoso,
  ``{Towards safe deep learning: accurately quantifying biomarker uncertainty
  in neural network predictions},'' in \emph{MICCAI}, 2018.

\bibitem{menze2014multimodal}
B.~H. Menze, A.~Jakab, S.~Bauer, J.~Kalpathy-Cramer, K.~Farahani, J.~Kirby,
  Y.~Burren, N.~Porz, J.~Slotboom, R.~Wiest \emph{et~al.}, ``The multimodal
  brain tumor image segmentation benchmark (brats),'' \emph{IEEE transactions
  on medical imaging}, vol.~34, no.~10, pp. 1993--2024, 2014.

\bibitem{Bragman2018}
F.~J.~S. Bragman, R.~Tanno, Z.~Eaton-Rosen, W.~Li, D.~J. Hawkes, S.~Ourselin,
  D.~C. Alexander, J.~R. McClelland, and M.~J. Cardoso, ``{Uncertainty in
  Multitask Learning: Joint Representations for Probabilistic MR-only
  Radiotherapy Planning},'' in \emph{MICCAI}, 2018, vol. 11073 LNCS, pp. 3--11.

\bibitem{Garg2018}
S.~Garg and S.~P. Awate, ``Perfect mcmc sampling in bayesian mrfs for
  uncertainty estimation in segmentation,'' in \emph{International Conference
  on Medical Image Computing and Computer-Assisted Intervention}.\hskip 1em
  plus 0.5em minus 0.4em\relax Springer, 2018, pp. 673--681.

\bibitem{Jungo2018}
A.~Jungo, R.~Meier, E.~Ermis, E.~Herrmann, and M.~Reyes, ``Uncertainty-driven
  sanity check: Application to postoperative brain tumor cavity segmentation,''
  \emph{arXiv preprint arXiv:1806.03106}, 2018.

\bibitem{Kwon2018}
Y.~Kwon, J.-H. Won, B.~J. Kim, and M.~C. Paik, ``Uncertainty quantification
  using bayesian neural networks in classification: Application to biomedical
  image segmentation,'' \emph{Computational Statistics \& Data Analysis}, vol.
  142, p. 106816, 2020.

\bibitem{winzeck2018isles}
S.~Winzeck, A.~Hakim, R.~McKinley, J.~A. Pinto, V.~Alves, C.~Silva, M.~Pisov,
  E.~Krivov, M.~Belyaev, M.~Monteiro \emph{et~al.}, ``Isles 2016 and
  2017-benchmarking ischemic stroke lesion outcome prediction based on
  multispectral mri,'' \emph{Frontiers in neurology}, vol.~9, p. 679, 2018.

\bibitem{Jungo2018_MICCAI}
A.~Jungo, R.~Meier, E.~Ermis, M.~Blatti-Moreno, E.~Herrmann, R.~Wiest, and
  M.~Reyes, ``On the effect of inter-observer variability for a reliable
  estimation of uncertainty of medical image segmentation,'' in
  \emph{International Conference on Medical Image Computing and
  Computer-Assisted Intervention}.\hskip 1em plus 0.5em minus 0.4em\relax
  Springer, 2018, pp. 682--690.

\bibitem{Soberanis-Mukul2019}
R.~D.~S. Mukul, N.~Navab, S.~Albarqouni \emph{et~al.}, ``An uncertainty-driven
  gcn refinement strategy for organ segmentation,'' \emph{Machine Learning for
  Biomedical Imaging}, vol.~1, no. MIDL 2020 special issue, pp. 1--10, 2020.

\bibitem{roth2015deeporgan}
H.~R. Roth, L.~Lu, A.~Farag, H.-C. Shin, J.~Liu, E.~B. Turkbey, and R.~M.
  Summers, ``Deeporgan: Multi-level deep convolutional networks for automated
  pancreas segmentation,'' in \emph{Medical Image Computing and
  Computer-Assisted Intervention--MICCAI 2015: 18th International Conference,
  Munich, Germany, October 5-9, 2015, Proceedings, Part I 18}.\hskip 1em plus
  0.5em minus 0.4em\relax Springer, 2015, pp. 556--564.

\bibitem{Seebock2019}
P.~Seeb{\"o}ck, J.~I. Orlando, T.~Schlegl, S.~M. Waldstein, H.~Bogunovi{\'c},
  S.~Klimscha, G.~Langs, and U.~Schmidt-Erfurth, ``Exploiting epistemic
  uncertainty of anatomy segmentation for anomaly detection in retinal oct,''
  \emph{IEEE transactions on medical imaging}, vol.~39, no.~1, pp. 87--98,
  2019.

\bibitem{Atrial}
A.~S.~C. dataset, \url{http://atriaseg2018.cardiacatlas.org/}.

\bibitem{armato2011lung}
S.~G. Armato~III, G.~McLennan, L.~Bidaut, M.~F. McNitt-Gray, C.~R. Meyer, A.~P.
  Reeves, B.~Zhao, D.~R. Aberle, C.~I. Henschke, E.~A. Hoffman \emph{et~al.},
  ``The lung image database consortium (lidc) and image database resource
  initiative (idri): a completed reference database of lung nodules on ct
  scans,'' \emph{Medical physics}, vol.~38, no.~2, pp. 915--931, 2011.

\bibitem{Awate2019}
S.~P. Awate, S.~Garg, and R.~Jena, ``Estimating uncertainty in mrf-based image
  segmentation: A perfect-mcmc approach,'' \emph{Medical image analysis},
  vol.~55, pp. 181--196, 2019.

\bibitem{wolterink2017automatic}
J.~M. Wolterink, T.~Leiner, M.~A. Viergever, and I.~I{\v{s}}gum, ``Automatic
  segmentation and disease classification using cardiac cine mr images,'' in
  \emph{International Workshop on Statistical Atlases and Computational Models
  of the Heart}.\hskip 1em plus 0.5em minus 0.4em\relax Springer, 2017, pp.
  101--110.

\bibitem{litjens2014computer}
G.~Litjens, O.~Debats, J.~Barentsz, N.~Karssemeijer, and H.~Huisman,
  ``Computer-aided detection of prostate cancer in mri,'' \emph{IEEE
  transactions on medical imaging}, vol.~33, no.~5, pp. 1083--1092, 2014.

\bibitem{litjens2014evaluation}
G.~Litjens, R.~Toth, W.~van~de Ven, C.~Hoeks, S.~Kerkstra, B.~van Ginneken,
  G.~Vincent, G.~Guillard, N.~Birbeck, J.~Zhang \emph{et~al.}, ``Evaluation of
  prostate segmentation algorithms for mri: the promise12 challenge,''
  \emph{Medical image analysis}, vol.~18, no.~2, pp. 359--373, 2014.

\bibitem{leclerc2019deep}
S.~Leclerc, E.~Smistad, J.~Pedrosa, A.~{\O}stvik, F.~Cervenansky, F.~Espinosa,
  T.~Espeland, E.~A.~R. Berg, P.-M. Jodoin, T.~Grenier \emph{et~al.}, ``Deep
  learning for segmentation using an open large-scale dataset in 2d
  echocardiography,'' \emph{IEEE transactions on medical imaging}, vol.~38,
  no.~9, pp. 2198--2210, 2019.

\bibitem{degerli2021early}
A.~Degerli, M.~Zabihi, S.~Kiranyaz, T.~Hamid, R.~Mazhar, R.~Hamila, and
  M.~Gabbouj, ``Early detection of myocardial infarction in low-quality
  echocardiography,'' \emph{IEEE Access}, vol.~9, pp. 34\,442--34\,453, 2021.

\bibitem{jaeger2014two}
S.~Jaeger, S.~Candemir, S.~Antani, Y.-X.~J. W{\'a}ng, P.-X. Lu, and G.~Thoma,
  ``Two public chest x-ray datasets for computer-aided screening of pulmonary
  diseases,'' \emph{Quantitative imaging in medicine and surgery}, vol.~4,
  no.~6, p. 475, 2014.

\bibitem{shiraishi2000development}
J.~Shiraishi, S.~Katsuragawa, J.~Ikezoe, T.~Matsumoto, T.~Kobayashi, K.-i.
  Komatsu, M.~Matsui, H.~Fujita, Y.~Kodera, and K.~doi, ``Development of a
  digital image database for chest radiographs with and without a lung nodule:
  receiver operating characteristic analysis of radiologists' detection of
  pulmonary nodules,'' \emph{American Journal of Roentgenology}, vol. 174,
  no.~1, pp. 71--74, 2000.

\bibitem{Luo2019}
J.~Luo, A.~Sedghi, K.~Popuri, D.~Cobzas, M.~Zhang, F.~Preiswerk, M.~Toews,
  A.~Golby, M.~Sugiyama, W.~M. Wells, and S.~Frisken, ``{On the Applicability
  of Registration Uncertainty},'' in \emph{MICCAI}, 2019, pp. 410--419.

\bibitem{LeFolgoc2017}
L.~{Le Folgoc}, H.~Delingette, A.~Criminisi, and N.~Ayache, ``{Quantifying
  Registration Uncertainty With Sparse Bayesian Modelling},'' \emph{IEEE
  Transactions on Medical Imaging}, vol.~36, no.~2, pp. 607--617, feb 2017.

\bibitem{Madsen2020}
D.~Madsen, A.~Morel-Forster, P.~Kahr, D.~Rahbani, T.~Vetter, and
  M.~L{\"{u}}thi, ``{A Closest Point Proposal for MCMC-based Probabilistic
  Surface Registration},'' in \emph{ECCV}, 2020, pp. 281--296.

\bibitem{Luo2020_miccai}
J.~Luo, S.~Frisken, D.~Wang, A.~Golby, M.~Sugiyama, and W.~{Wells III}, ``{Are
  Registration Uncertainty and Error Monotonically Associated?}'' in
  \emph{MICCAI}, 2020, pp. 264--274.

\bibitem{Markiewicz2021}
P.~J. Markiewicz, J.~C. Matthews, J.~Ashburner, D.~M. Cash, D.~L. Thomas,
  E.~{De Vita}, A.~Barnes, M.~J. Cardoso, M.~Modat, R.~Brown, K.~Thielemans,
  C.~da~Costa-Luis, I.~{Lopes Alves}, J.~D. Gispert, M.~E. Schmidt, P.~Marsden,
  A.~Hammers, S.~Ourselin, and F.~Barkhof, ``{Uncertainty analysis of MR-PET
  image registration for precision neuro-PET imaging},'' \emph{NeuroImage},
  vol. 232, p. 117821, may 2021.

\bibitem{Xu2022_miccai}
Z.~Xu, J.~Luo, D.~Lu, J.~Yan, S.~Frisken, J.~Jagadeesan, W.~M. Wells, X.~Li,
  Y.~Zheng, and R.~K.-y. Tong, ``{Double-Uncertainty Guided Spatial and
  Temporal Consistency Regularization Weighting for Learning-Based Abdominal
  Registration},'' in \emph{MICCAI}, 2022, pp. 14--24.

\bibitem{Zhang2019_cvpr}
Z.~Zhang, A.~Romero, M.~J. Muckley, P.~Vincent, L.~Yang, and M.~Drozdzal,
  ``{Reducing Uncertainty in Undersampled MRI Reconstruction with Active
  Acquisition},'' in \emph{CVPR}, feb 2019.

\bibitem{Edupuganti2021}
V.~Edupuganti, M.~Mardani, S.~Vasanawala, and J.~Pauly, ``{Uncertainty
  Quantification in Deep MRI Reconstruction},'' \emph{IEEE Transactions on
  Medical Imaging}, vol.~40, no.~1, pp. 239--250, jan 2021.

\bibitem{Sudarshan2021}
V.~P. Sudarshan, U.~Upadhyay, G.~F. Egan, Z.~Chen, and S.~P. Awate, ``{Towards
  lower-dose PET using physics-based uncertainty-aware multimodal learning with
  robustness to out-of-distribution data},'' \emph{Medical Image Analysis},
  vol.~73, p. 102187, oct 2021.

\bibitem{Armanious2021}
K.~Armanious, S.~Abdulatif, W.~Shi, T.~Hepp, S.~Gatidis, and B.~Yang,
  ``{Uncertainty-Based Biological Age Estimation of Brain MRI Scans},'' in
  \emph{ICASSP}.\hskip 1em plus 0.5em minus 0.4em\relax IEEE, jun 2021, pp.
  1100--1104.

\bibitem{Narnhofer2022}
D.~Narnhofer, A.~Effland, E.~Kobler, K.~Hammernik, F.~Knoll, and T.~Pock,
  ``{Bayesian Uncertainty Estimation of Learned Variational MRI
  Reconstruction},'' \emph{IEEE Transactions on Medical Imaging}, vol.~41,
  no.~2, pp. 279--291, feb 2022.

\bibitem{Cui2022_miccai}
J.~Cui, Y.~Xie, A.~A. Joshi, K.~Gong, K.~Kim, Y.-D. Son, J.-H. Kim, R.~Leahy,
  H.~Liu, and Q.~Li, ``{PET Denoising and Uncertainty Estimation Based on NVAE
  Model Using Quantile Regression Loss},'' pp. 173--183, 2022.

\bibitem{Tanno2017}
R.~Tanno, D.~E. Worrall, A.~Ghosh, E.~Kaden, S.~N. Sotiropoulos, A.~Criminisi,
  and D.~C. Alexander, ``Bayesian image quality transfer with cnns: exploring
  uncertainty in dmri super-resolution,'' in \emph{International Conference on
  Medical Image Computing and Computer-Assisted Intervention}.\hskip 1em plus
  0.5em minus 0.4em\relax Springer, 2017, pp. 611--619.

\bibitem{Tanno2019}
R.~Tanno, D.~Worrall, E.~Kaden, A.~Ghosh, F.~Grussu, A.~Bizzi, S.~N.
  Sotiropoulos, A.~Criminisi, and D.~C. Alexander, ``Uncertainty quantification
  in deep learning for safer neuroimage enhancement,'' \emph{arXiv preprint
  arXiv:1907.13418}, 2019.

\bibitem{Tanno2021}
R.~Tanno, D.~E. Worrall, E.~Kaden, A.~Ghosh, F.~Grussu, A.~Bizzi, S.~N.
  Sotiropoulos, A.~Criminisi, and D.~C. Alexander, ``{Uncertainty modelling in
  deep learning for safer neuroimage enhancement: Demonstration in diffusion
  MRI},'' \emph{NeuroImage}, vol. 225, p. 117366, jan 2021.

\bibitem{eaton2019easy}
Z.~Eaton-Rosen, T.~Varsavsky, S.~Ourselin, and M.~J. Cardoso, ``As easy as 1,
  2... 4? uncertainty in counting tasks for medical imaging,'' in
  \emph{International Conference on Medical Image Computing and
  Computer-Assisted Intervention}.\hskip 1em plus 0.5em minus 0.4em\relax
  Springer, 2019, pp. 356--364.

\bibitem{Ozdemir2017}
O.~Ozdemir, B.~Woodward, and A.~A. Berlin, ``{Propagating Uncertainty in
  Multi-Stage Bayesian Convolutional Neural Networks with Application to
  Pulmonary Nodule Detection},'' \emph{arXiv}, 2017.

\bibitem{Petersen2019}
J.~Petersen, P.~F. J{\"a}ger, F.~Isensee, S.~A. Kohl, U.~Neuberger, W.~Wick,
  J.~Debus, S.~Heiland, M.~Bendszus, P.~Kickingereder \emph{et~al.}, ``Deep
  probabilistic modeling of glioma growth,'' in \emph{International Conference
  on Medical Image Computing and Computer-Assisted Intervention}.\hskip 1em
  plus 0.5em minus 0.4em\relax Springer, 2019, pp. 806--814.

\bibitem{CUMC12}
CUMC12, \url{https://www.synapse.org/\#!Synapse:syn3207203}.

\bibitem{van2013wu}
D.~C. Van~Essen, S.~M. Smith, D.~M. Barch, T.~E. Behrens, E.~Yacoub,
  K.~Ugurbil, W.-M.~H. Consortium \emph{et~al.}, ``The wu-minn human connectome
  project: an overview,'' \emph{Neuroimage}, vol.~80, pp. 62--79, 2013.

\bibitem{lifespan}
Lifespan, \url{http://lifespan.humanconnectome.org}.

\bibitem{alexander2017image}
D.~C. Alexander, D.~Zikic, A.~Ghosh, R.~Tanno, V.~Wottschel, J.~Zhang,
  E.~Kaden, T.~B. Dyrby, S.~N. Sotiropoulos, H.~Zhang \emph{et~al.}, ``Image
  quality transfer and applications in diffusion mri,'' \emph{NeuroImage}, vol.
  152, pp. 283--298, 2017.

\bibitem{figini2018prediction}
M.~Figini, M.~Riva, M.~Graham, G.~M. Castelli, B.~Fernandes, M.~Grimaldi,
  G.~Baselli, F.~Pessina, L.~Bello, H.~Zhang \emph{et~al.}, ``Prediction of
  isocitrate dehydrogenase genotype in brain gliomas with mri: single-shell
  versus multishell diffusion models,'' \emph{Radiology}, vol. 289, no.~3, pp.
  788--796, 2018.

\bibitem{setio2017validation}
A.~A.~A. Setio, A.~Traverso, T.~De~Bel, M.~S. Berens, C.~Van Den~Bogaard,
  P.~Cerello, H.~Chen, Q.~Dou, M.~E. Fantacci, B.~Geurts \emph{et~al.},
  ``Validation, comparison, and combination of algorithms for automatic
  detection of pulmonary nodules in computed tomography images: the luna16
  challenge,'' \emph{Medical image analysis}, vol.~42, pp. 1--13, 2017.

\bibitem{naylor2018segmentation}
P.~Naylor, M.~La{\'e}, F.~Reyal, and T.~Walter, ``Segmentation of nuclei in
  histopathology images by deep regression of the distance map,'' \emph{IEEE
  transactions on medical imaging}, vol.~38, no.~2, pp. 448--459, 2018.

\bibitem{kuijf2019standardized}
H.~J. Kuijf, J.~M. Biesbroek, J.~De~Bresser, R.~Heinen, S.~Andermatt, M.~Bento,
  M.~Berseth, M.~Belyaev, M.~J. Cardoso, A.~Casamitjana \emph{et~al.},
  ``Standardized assessment of automatic segmentation of white matter
  hyperintensities and results of the wmh segmentation challenge,'' \emph{IEEE
  transactions on medical imaging}, vol.~38, no.~11, pp. 2556--2568, 2019.

\bibitem{zbontar2018fastmri}
J.~Zbontar, F.~Knoll, A.~Sriram, T.~Murrell, Z.~Huang, M.~J. Muckley,
  A.~Defazio, R.~Stern, P.~Johnson, M.~Bruno \emph{et~al.}, ``fastmri: An open
  dataset and benchmarks for accelerated mri,'' \emph{arXiv preprint
  arXiv:1811.08839}, 2018.

\bibitem{deng2009imagenet}
J.~Deng, W.~Dong, R.~Socher, L.-J. Li, K.~Li, and L.~Fei-Fei, ``Imagenet: A
  large-scale hierarchical image database,'' in \emph{2009 IEEE conference on
  computer vision and pattern recognition}.\hskip 1em plus 0.5em minus
  0.4em\relax Ieee, 2009, pp. 248--255.

\bibitem{gerig2018morphable}
T.~Gerig, A.~Morel-Forster, C.~Blumer, B.~Egger, M.~Luthi, S.~Sch{\"o}nborn,
  and T.~Vetter, ``Morphable face models-an open framework,'' in \emph{2018
  13th IEEE International Conference on Automatic Face \& Gesture Recognition
  (FG 2018)}.\hskip 1em plus 0.5em minus 0.4em\relax IEEE, 2018, pp. 75--82.

\bibitem{xiao2017resect}
Y.~Xiao, M.~Fortin, G.~Unsg{\aa}rd, H.~Rivaz, and I.~Reinertsen, ``Resect: a
  clinical database of pre-operative mri and intra-operative ultrasound in
  low-grade glioma surgeries,'' \emph{Med. Phys}, vol.~44, no.~7, pp.
  3875--3882, 2017.

\bibitem{mridata}
Mridata.org, \url{http://mridata.org/}.

\bibitem{lamontagne2019oasis}
P.~J. LaMontagne, T.~L. Benzinger, J.~C. Morris, S.~Keefe, R.~Hornbeck,
  C.~Xiong, E.~Grant, J.~Hassenstab, K.~Moulder, A.~G. Vlassenko \emph{et~al.},
  ``Oasis-3: longitudinal neuroimaging, clinical, and cognitive dataset for
  normal aging and alzheimer disease,'' \emph{MedRxiv}, pp. 2019--12, 2019.

\bibitem{ginovart200311c}
N.~Ginovart, A.~A. Wilson, J.~H. Meyer, D.~Hussey, and S.~Houle, ``[11c]-dasb,
  a tool for in vivo measurement of ssri-induced occupancy of the serotonin
  transporter: Pet characterization and evaluation in cats,'' \emph{Synapse},
  vol.~47, no.~2, pp. 123--133, 2003.

\bibitem{knoll2020fastmri}
F.~Knoll, J.~Zbontar, A.~Sriram, M.~J. Muckley, M.~Bruno, A.~Defazio,
  M.~Parente, K.~J. Geras, J.~Katsnelson, H.~Chandarana \emph{et~al.},
  ``fastmri: A publicly available raw k-space and dicom dataset of knee images
  for accelerated mr image reconstruction using machine learning,''
  \emph{Radiology: Artificial Intelligence}, vol.~2, no.~1, p. e190007, 2020.

\end{thebibliography}

\end{document}